\documentclass[%
 reprint,
 amsmath,amssymb,
 aps,
]{revtex4-2}

\usepackage{amsmath}

\usepackage{xcolor}
\usepackage{graphicx}
\usepackage{dcolumn}
\usepackage{bm}

\begin{document}

\preprint{APS/123-QED}

\title{Measuring hierarchically-organized interactions in dynamic networks through spectral entropy rates: theory, estimation,  and illustrative application \\ to physiological networks} 


\author{Laura Sparacino}%
 \altaffiliation{Department of Engineering, University of Palermo, Italy}
 
\author{Yuri Antonacci}
\altaffiliation{Department of Engineering, University of Palermo, Italy}

\author{Gorana Mijatovic}
 \altaffiliation{Faculty of Technical Sciences, University of Novi Sad, Serbia}
 \email{gorana86@uns.ac.rs}

\author{Luca Faes}
\altaffiliation{Department of Engineering, University of Palermo, Italy}
\email{luca.faes@unipa.it}

\date{\today}

\begin{abstract}
Recent advances in signal processing and information theory are boosting the development of new approaches for the data-driven modelling of complex network systems. In the fields of Network Physiology and Network Neuroscience where the signals of interest are often rich of oscillatory content, the spectral representation of network systems is essential to ascribe the analyzed interactions to specific oscillations with physiological meaning.
In this context, the present work formalizes a coherent framework which integrates several information dynamics approaches to quantify node-specific, pairwise and higher-order interactions in network systems. The framework establishes a hierarchical organization of interactions of different order using measures of entropy rate, mutual information rate and O-information rate, to quantify respectively the dynamics of individual nodes, the links between pairs of nodes, and the redundant/synergistic hyperlinks between groups of nodes. All measures are formulated in the time domain, and then expanded to the spectral domain to obtain frequency-specific information.
The practical computation of all measures is favored presenting a toolbox that implements their parametric and non-parametric estimation, and includes approaches to assess their statistical significance. The framework is illustrated first using theoretical examples where the properties of the measures are displayed in benchmark simulated network systems, and then applied to representative examples of multivariate time series in the context of Network Neuroscience and Network Physiology.
\end{abstract}

\maketitle

\section{INTRODUCTION}
The increasing availability of large-scale and fine-grained recordings of biomedical signals is opening the way to the network representation of complex physiological systems. For instance, in neuroscience the organizational principles of functional segregation and integration in the human brain are typically studied through the theoretical and empirical tools of Network Neuroscience \citep{bassett2017network}, while in integrative physiology the reductionist approach of studying in isolation the function of an organ system is nowadays complemented by the holistic investigation of collective interactions among diverse organ systems performed in the field of Network Physiology \citep{bashan2012network}. Network Neuroscience and Network Physiology are subfields of Network Science, a large interdisciplinary area that develops theoretical and practical techniques to improve the understanding of natural and man-made networks with hierarchical structures \citep{barabasi2013network}.

Data-driven methods for network inference play a key role in Network Neuroscience and Network Physiology, being devised to build a network model out of a set of observed multivariate time series. Such model is typically encoded by a graph where the observed dynamical system (e.g., the brain or the human organism) is represented by distinct nodes (e.g., neural units or organ systems) connected by edges mapping functional dependencies (e.g., brain connectivity or cardiovascular interactions) \citep{rubinov2010complex,lehnertz2020human}. Beyond this basic description, the need to deepen the exploration of real-world systems has led network scientists to enrich the way to represent the system properties captured by a network model \citep{butts2009revisiting}. Augmented network descriptions have been proposed which make use, for instance, of active nodes encoding self-dependencies within an individual process, and of directed and/or weighted edges depicting cause-effect relations and quantifying the intensity of interactions. These representations are well accommodated in functional brain and physiological networks through the definition of measures to assess complexity or regularity of individual time series \citep{pincus1994physiological}, or coupling and causality between pairs of time series \citep{pereda2005nonlinear}.

In spite of the usefulness of network models encoded by graphs, the representation with self-effects and pairwise (dyadic) interactions is often insufficient to provide a complete description of a complex system. It is now firmly acknowledged that many real-world systems display high-order (polyadic) interactions, i.e. interactions involving more than two network nodes \citep{battiston2020networks}. Thus, in these systems the network behavior is integrated at different hierarchical levels and time scales. This occurs also in Network Neuroscience and Network Physiology, where it is important to distinguish between brain regions or organ systems that interact as a pair, or as a part of a more complex structure, to produce the observed dynamics. For instance, brain dynamics display mesoscopic or macroscopic behaviors requiring multiple-unit interactions to be predicted accurately \citep{stramaglia2014synergy}, and cardiovascular interactions may arise autonomously from self-sustained mechanisms or as a result of the effects of respiration on the measured dynamics \citep{faes2016information}. The generalized network structure which allows to go beyond the framework of pairwise interactions is the so-called "hierarchical high-order network" (hHON), which is described by mathematical constructs such as simplicial complexes and hypergraphs \citep{courtney2016generalized}. This novel representation is impacting strongly on the ability to describe the real-world systems studied in the context of Network Science \citep{battiston2020networks}.

\begin{figure*}[tbp]
    \centering \includegraphics[scale=1.05] {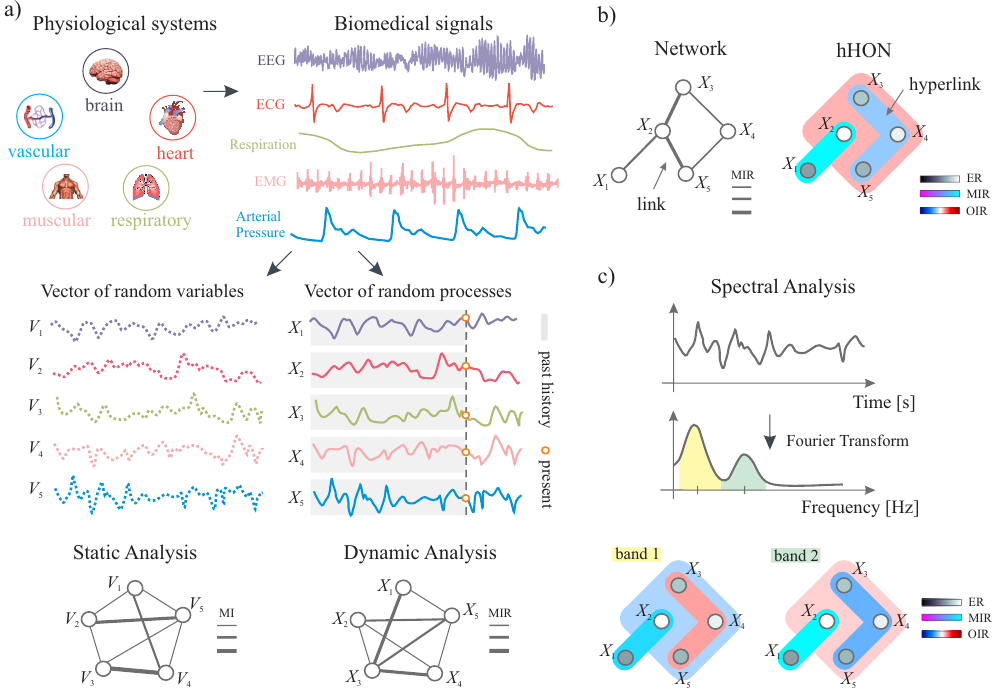}
    \caption{\textbf{Information-theoretic and spectral representations of hierarchically-organized interactions in network systems.}
    \textbf{a)} In Network Physiology, collective interactions among diverse organ systems (e.g., brain, heart, vascular, muscular and respiratory systems) are investigated recording biosignals like the EEG, ECG, arterial pressure, EMG and respiration, from which time series representing the dynamical activity of these systems are extracted. The data collected in these series can be considered as a a realization of a vector of random variables ($V_i$) or random processes ($X_i$). 
    The network analysis of this data is \textit{static} when measures like the mutual information (MI) are used to connect pairs of random variables, and \textit{dynamic} when measures like the MI rate (MIR) are used to connect pairs of random vectors considering their temporal correlations.
    \textbf{b)} The multiple interactions in the physiological network can be investigated in different ways: traditional multivariate analysis is still anchored to the concept of interaction between two nodes even though the other nodes are taken into account, thus yielding the standard network representation where the link is the building block and is assessed by pairwise measures such as the MIR; the analysis of higher-order interactions moves forward, encoding such interactions by the hyperlinks of a hierarchical high-order network (hHON) where interactions of different order are represented using different dynamic information measures (e.g., order 1: entropy rate (ER); order 2: MIR; order $>$2: O-information rate (OIR)).
    \textbf{c)} In physiological networks with oscillatory node activity, the shift from the time to the frequency domain representation  is essential to capture the wide range of time scales characterizing the dynamical activity at the nodes; in this case, information measures can be expanded in the spectral domain to obtain frequency-specific information on the hHON interactions occurring within distinct frequency bands.  
    }
    \label{fig:intro}
\end{figure*}

While the implementation of hHONs is straightforward for networks inherently defined as sets of interactions, it is much less striking in systems where interactions are not already identified but need to be inferred from data, as in the case of brain and physiological networks. The main reason for this difficulty is that measures to quantify polyadic interactions from time series data have not been defined unequivocally \citep{wibral2014directed,james2016information,faes2016information,lizier2018information}.
Different information-theoretic frameworks performing entropy decomposition of the multiple time series mapping the activity of network systems provide tools, such as the measures of redundant and synergistic information shared by groups of source time series about a target series, akin to the detection of higher-order effects \citep{faes2016information,lizier2018information}. Thus, these frameworks could be exploited for the detection and estimation of high-order interactions in practical settings. In this context, the unification and the extension to high-order interactions of emerging approaches to treat diverse types of physiological activity, as well as the computation of quantities able to reflect how physiological oscillations are deployed across several time scales \citep{faes2016information,faes2021information}, performed via reliable estimators from collected data, would open new perspectives for the use of hHON structures in Network Neuroscience and Network Physiology.

The aim of this tutorial paper is to integrate, extend, unify and illustrate thoroughly several classic and recent approaches proposed for the analysis of network systems which are gaining wide interest in the fields of Network Neuroscience and Physiology (\cite{mcgill1954multivariate, gelfand1959calculation, kolmogorov1959entropy, duncan1970calculation, cover1999elements, chicharro2011spectral, rosas2019quantifying, faes2021information, antonacci2021measuring, faes2022new, bara2023comparison}).
These approaches are based on the information-theoretic and spectral representations of interactions in network systems, and are here presented along three different lines of development (Fig. \ref{fig:intro}).
First, we move from the standard \textit{static analysis} of physiological processes, which draws a parallel between physiological networks and vectors of random variables, to a \textit{dynamic analysis} which models the observed network system in terms of vector random processes (Fig. \ref{fig:intro}a); this is achieved moving from the use of measures of entropy computed for random variables to the use of measures of entropy rate which explicitly consider the temporal correlations within and between random processes. 
Second, when we consider the interactions in a network, we shift from the paradigm of \textit{multivariate analysis} to the paradigm of \textit{higher-order interactions}; while multivariate analysis
is focused on the activity of two nodes of the network even when the other nodes are taken into account, measures of higher-order interaction focus on more than two network nodes providing an overall quantification of their collective interaction (see Fig. \ref{fig:intro}b, where multivariate and higher-order measures are encoded by links in classical networks and hyperlinks in high-order networks).
Third, when we study networks where the activity at the nodes is rich of oscillatory content, we shift from the \textit{time domain} to the \textit{frequency domain} representation of the network, where information measures are expanded in the spectral domain to obtain frequency-specific information (Fig. \ref{fig:intro}c).

In the paper, the approaches for the analysis of interactions in network systems are unified in a coherent framework where spectral measures of entropy rate are linked with their time-domain analogous measures, and are categorized hierarchically  on the basis of the number of network nodes involved in the computation of each measure.
After providing rigorous theoretical definitions (Sect. \ref{sec_methods}), particular emphasis is put on the practical computation of the measures from time series data, presenting both parametric and non-parametric estimators as well as approaches to assess the statistical significance of the estimated measures (Sect. \ref{practical_implementation}). The framework is illustrated first using theoretical examples where the properties of each measure are demonstrated in benchmark simulated network systems (Sect. \ref{th_simulations}), and then applied to paradigmatic examples of multivariate time series (Sect. \ref{applications}) in the context of Network Neuroscience (multichannel electroencephalographic signals) and Network Physiology (heart rate, arterial pressure, respiration and cerebral blood flow velocity time series). 



\section{Framework for the analysis of interactions in Network Systems} \label{sec_methods}
This section formalizes the framework to measure time domain and frequency-specific interactions of different order in networks of multiple interconnected systems. The framework is defined for \textit{dynamic systems}, i.e. systems evolving over time whose activity is modeled appropriately in the context of multivariate random processes. Nevertheless, since the information-theoretic analysis of random processes is performed using information measures applied to random variables, we will start reviewing the entropy measures adopted to describe static networks mapped by random variables (Sect. \ref{StaticAnalysis}). Then, these measures are extended to the analysis of interactions in dynamic networks mapped by random processes exploiting the concept of entropy rate  (Sect. \ref{DynamicAnalysis}), and formalized for linear processes studied in the frequency domain using spectral entropy rates (Sect. \ref{SpectralAnalysis}).


\subsection{Static networks of random variables} \label{StaticAnalysis}
\begin{figure*} 
    \centering \includegraphics[scale=1] {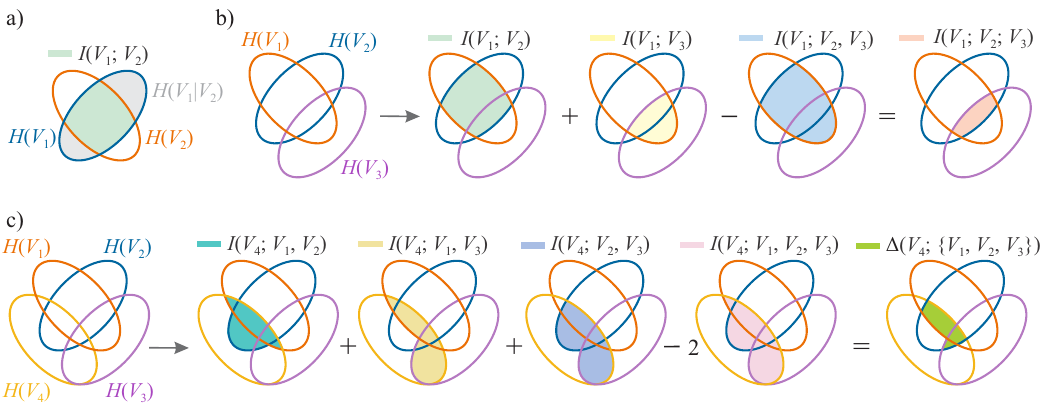}
    \caption{\textbf{Venn-diagram representation of the information measures quantifying hierarchically-organized interactions in static networks of random variables.}    
    \textbf{a)} Mutual information between two random variables $V_1$ and $V_2$, $I(V_1; V_2)$, obtained as the difference between the entropy $H(V_1)$ and the conditional entropy $H(V_1|V_2)$. \textbf{b)} Interaction information among the three variables $V_1$, $V_2$ and $V_3$, $I(V_1; V_2; V_3)$, obtained according to Eq. (\ref{InteractionInformation}). \textbf{c)} Gradient of the O-information, $\Delta (V_4; \{V_1, V_2, V_3\})$, quantifying the information increment obtained when a fourth variable $V_4$ is added to the group of random variables $\{V_1, V_2, V_3\}$, obtained according to Eq. (\ref{DeltaOI}); the O-information among these four variables is obtained summing the interaction information $I(V_1; V_2; V_3)$ and the gradient $\Delta (V_4; \{V_1, V_2, V_3\})$ (see Eq. (\ref{OInformation})).}
    \label{fig:venn_var}
\end{figure*}

Let us consider a static network system $\mathcal{V}$ composed of $M$ nodes $\mathcal{V}_1,\ldots,\mathcal{V}_M$, where the activity at each node is described by the (possibly vector) random variable $V_i, i=1,\ldots M$. By using an information-theoretic perspective, the activity at the $i^{\mathrm{th}}$ node of the network can be assessed through the information carried by the variable $V_i$, which is quantified by the well-known entropy measure (\cite{shannon1948mathematical, cover1999elements}):
\begin{equation}
   H(V_i)=\mathbb{E} \left[ \ln \frac{1}{p(v_i)} \right],
   \label{Entropy}
\end{equation}
while the interaction between the activities of the $i^{\mathrm{th}}$ and $j^{\mathrm{th}}$ nodes can be assessed through the mutual information (MI) between $V_i$ and $V_j$:
\begin{equation}
   I(V_i;V_j)=\mathbb{E} \left[ \ln \frac{p(v_i,v_j)}{p(v_i)p(v_j)} \right],
   \label{MutualInformation}
\end{equation}
where $p(\cdot,\cdot)$ and $p(\cdot)$ denote joint and marginal probability distribution, and $\mathbb{E}[\cdot]$ is the expectation operator. The entropy quantifies the information contained in a random variable intended as the average uncertainty about its outcomes, while the MI quantifies the information shared by two variables intended as the uncertainty about one variable that is resolved by knowing the other variable. Remarkably, the MI is a symmetric measure (i.e., $I(V_i;V_j)=I(V_j;V_i)$), and is linked to the joint and individual entropies of the two variables by the relation $I(V_i;V_j)=H(V_i)+H(V_j)-H(V_i,V_j)$. The latter can be also expressed as $I(V_i;V_j)=H(V_i)-H(V_i|V_j)$, where $H(V_i|V_j)=H(V_i,V_j)-H(V_j)$ is the conditional entropy (CE) of $V_i$ given $V_j$, quantifying the information carried by one variable that is not shared with the other, intended as the residual uncertainty which remains in one variable when the other variable is known. The relations between entropy, MI and CE are depicted in the Venn diagram representation in Fig. \ref{fig:venn_var}a.

Then, the interaction among three variables $V_i$, $V_j$ and $V_k$, $i,j,k \in \{1,\ldots\ M\}$, is quantified by the interaction information (II) (\cite{mcgill1954multivariate}), which compares the information that one target variable, say $V_i$, shares with two source variables, say $V_j$ and $V_k$, when the sources are taken separately but not when they are taken together. Accordingly, the II is quantified subtracting the MI between the target and the two sources from the sum of the MIs between the target and each source:
\begin{equation}
   I(V_i;V_j;V_k)=I(V_i;V_j)+I(V_i;V_k)-I(V_i;V_j,V_k).
   \label{InteractionInformation}
\end{equation}
The computation of the II is illustrated using Venn diagrams in Fig. \ref{fig:venn_var}b. Importantly, the II is symmetric (i.e., it does not change modifying the target variable) and can take either positive or negative values. Specifically, the II is positive if the two sources share more information with the target when they are considered individually, denoting \textit{redundancy}; on the contrary, the II is negative if the two sources share more information with the target when they are considered jointly, denoting \textit{synergy}.

The II has been recently generalized to allow the information-theoretic analysis of high order interactions among an arbitrarily large number of random variables through the definition of the so-called O-information (OI) (\cite{rosas2019quantifying}). The OI among $N$ variables can be defined as the sum of the OI for a subset including $N-1$ variables plus a \textit{gradient} quantifying the increment obtained when a new variable is added. Specifically, when the variable $V_N$ is added to the set $V^{N-1}=\{V_1,\ldots,V_{N-1}\}$ to form the set $V^N=\{V_1,\ldots,V_N\}$, the OI can be formulated as
\begin{equation}
   \Omega(V^N)=\Omega(V^{N-1}) + \Delta(V_N;V^{N-1}),
   \label{OInformation}
\end{equation}
where the gradient is defined as
\begin{equation}
   \Delta(V_N;V^{N-1})=\sum_{i=1}^{N-1}I(V_N;V^{N-1}_{-i})+(2-N)I(V_N;V^{N-1}),
   \label{DeltaOI}
\end{equation}
with $V^{N-1}_{-i}=V^{N-1}\backslash V_i$. The computation of the gradient for the case of $N=4$ variables is illustrated using Venn diagrams in Fig. \ref{fig:venn_var}c. Crucially, since the OI for two random variables is null ($\Omega(V^2)=0$), the OI for three variables is equal to the gradient, which in turn corresponds to the II: $\Omega(V^3)=\Delta(V_3;V^2)=I(V_3;V_2)+I(V_3;V_1)-I(V_3;V_1,V_2)=I(V_1;V_2;V_3)$; these formulations allow the iterative computation of the OI through progressive inclusion of variables (\cite{rosas2019quantifying}).
Importantly, both the gradient and the O-information can be positive or negative, reflecting respectively redundant and synergistic high-order interactions in the analyzed set of random variables.

\subsection{Dynamic networks of random processes: time domain analysis} \label{DynamicAnalysis}

Let us consider a dynamic network system $\mathcal{X}$ composed of $M$ nodes $\mathcal{X}_1,\ldots,\mathcal{X}_M$, where the activity at each node is described in terms of random processes. Specifically, we consider  $Q$ stationary stochastic processes $Y=\{Y_1, \ldots, Y_Q\}$, grouped in $M$ blocks $X=\{X_1, \ldots, X_M\}$, and assume that each block process  $X_i=\{X_{i_1},\ldots,X_{i_{M_i}} \}$ describes the dynamic activity of the network node $\mathcal{X}_i, i=1,\ldots,M$; the $i^{\mathrm{th}}$ block has dimension $M_i$, so that $Q=\sum_{i=1}^M M_i$. Then, to highlight the dynamic nature of the process $X_i$, we denote as $X_{i,n}$, $X^k_{i,n}=[X_{i,n-1} \cdots X_{i,n-k}]$, and $X^-_{i,n}=\lim_{k \to \infty}X^k_{i,n}$ the random variables that sample the process  at the present time $n$, over the past $k$ lags, and over the whole past history, respectively.

\begin{figure*} [tbp]
    \centering \includegraphics[scale=1] {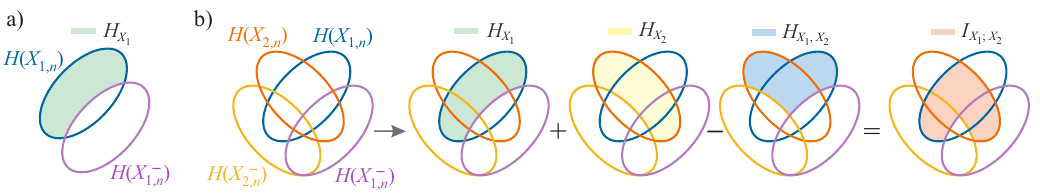}
    \caption{\textbf{Venn-diagram representation of the information measures quantifying hierarchically-organized interactions in dynamic networks of random processes.}
    \textbf{a)} Entropy rate of the process $X_1$, $H_{X_1}$, obtained as the conditional entropy of the variable representing the present state of the process $X_{1,n}$ given the variables sampling its past history  $X_{1,n}^-$. \textbf{b)} Mutual information rate between two processes $X_1$ and $X_2$, $I_{X_1; X_2}$, obtained according to Eq. (\ref{MIR_HR}). The MIR is the basic measure used for the computation of higher-order measures like the IIR of three processes (Eq. (\ref{IIR})) or the OIR of more than three processes (Eq. (\ref{OIR})).}
    \label{fig:venn_proc}
\end{figure*}

With the above notation, considering the statistical dependencies between the present and past states of the analyzed systems, it is possible to define dynamic information measures which extend to random processes the measures defined in Sect. \ref{StaticAnalysis} for random variables. The information-theoretic analysis of dynamic information exploits the concept of \textit{entropy rate} (ER), which quantifies the rate of generation of new information in a random process. Specifically, for the stationary random process $X_i$, the ER is defined using these two equivalent definitions (\cite{cover1999elements, chicharro2011spectral}):
\begin{equation}
   H_{X_i}=\lim_{m \to \infty}\frac{1}{m} H(X_{i,n:n+m})=H(X_{i,n}|X^-_{i,n}),
   \label{ER}
\end{equation}
where the second formulation evidences the conditional entropy of the variable representing the present of the process given the variables sampling its past history. Therefore, the ER reflects the complexity of the process intended as the unpredictability of its present state given the past, ranging from $H_{X_i}=0$ for a completely self-predictable process to $H_{X_i}=H(X_{i,n})$ for a fully unpredictable process without temporal statistical structure. In the framework for the analysis of dynamic interactions presented in this work, the ER quantifies the so-called interactions of order one, i.e. the interactions occurring internally in the $i^{\mathrm{th}}$ node of the analyzed network which is mapped by the process $X_i$. The Venn diagram illustration of the ER is given in Fig. \ref{fig:venn_proc}a.

The interactions of order two (i.e., pairwise interactions) between the $i^{\mathrm{th}}$  and $j^{\mathrm{th}}$  nodes of the analyzed dynamic network are assessed by the the \textit{mutual information rate} (MIR) computed between the processes $X_i$ and $X_j$. The MIR quantifies the information shared by the two processes per unit of time (\cite{duncan1970calculation}):
\begin{equation}
   I_{X_i;X_j}=\lim_{m \to \infty}\frac{1}{m} I(X_{i,n:n+m};X_{j,n:n+m}).
   \label{MIR}
\end{equation}
It is a symmetric measure (i.e., $I_{X_i;X_j}=I_{X_j;X_i}$), and can be obtained from the present and past information of the processes by summing the entropy rate of the first process to the entropy rate of the second process, and then subtracting the joint entropy rate of the two processes (see also Fig. \ref{fig:venn_proc}b): 
\begin{equation}
   I_{X_i;X_j}=H_{X_i}+H_{X_j}-H_{X_i,X_j}.
   \label{MIR_HR}
\end{equation}

The MIR is a measure of dynamic coupling, and can also be used as a building block for the assessment of higher-order interactions in random processes. Specifically, the dynamic interaction of order three among the processes $X_i$, $X_j$ and $X_k$ can be quantified by the \textit{interaction information rate} (IIR) (\cite{faes2021information}) by using MIR terms in a formulation similar to (\ref{InteractionInformation}):
\begin{equation}
I_{X_i;X_j;X_k}=I_{X_i;X_j}+I_{X_i;X_k}-I_{X_i;X_j,X_k}.
   \label{IIR}
\end{equation}

Moreover, exploiting the same expressions valid for the OI (Eqs. (\ref{OInformation}), (\ref{DeltaOI})) where the MIR of random processes is used in place of the MI of random variables, it is possible to define a so-called \textit{O-information rate} (OIR) which can be computed iteratively using gradients. Specifically, the OIR of a group of random processes $X^N=\{X_1,\ldots,X_{N}\}$ is computed from the OIR of a subset including $N-1$ processes, e.g.,  $X^{N-1}=\{X_1,\ldots,X_{N-1}\}$, summing a gradient which quantities the increment obtained when the process $X_N$ is appended to $X^{N-1}$ (\cite{faes2022new}):
\begin{align}
   \Omega_{X^N} &=\Omega_{X^{N-1}} + \Delta_{X_N;X^{N-1}}, \label{OIR} \\
   \Delta_{X_N;X^{N-1}} &=\sum_{i=1}^{N-1}I_{X_N;X^{N-1}_{-i}}+(2-N)I_{X_N;X^{N-1}},
   \label{DeltaOIR}
\end{align}
with $X^{N-1}_{-i}=X^{N-1}\backslash X_i$.
According to this definition, the OIR is zero for any two processes ($\Omega_{X_i,X_j}=0$), and is equal to the gradient for three processes, i.e. $\Omega_{X_i,X_j,X_k}=\Delta_{X_i;X_j,X_k}$, which in turns corresponds to the IIR (\ref{IIR}) with $N=3$. Both the IIR of three processes and the OIR computed for $N>3$ processes are symmetric, i.e. do not change if the order of the processes is swapped in the computation. Importantly, these measures, as well as the gradients, can be either positive or negative, with the sign reflecting the redundant or synergistic nature of the interactions in groups of random processes; specifically, positive values of $\Omega_{X^N}$ or $\Delta_{X_N;X^{N-1}}$ denote redundancy, while negative values denote synergy. 

Remarkably, combinatorial explosion in the context of high-order interactions in physiological networks, which refers to the rapid increase in the number of possible node combinations as the size of the network grows, is an open challenge in the field of Network Science. In a network with $M$ nodes, the activity of the single entities is described by means of $M$ values of ER, and $\frac{M (M-1)}{2}$ values of MIR. However, when we start looking at interactions beyond just pairs of nodes (i.e., high-order interactions), the complexity skyrockets. In general, for the $M$-node network, the number of high-order interactions assessed via the OIR measure scales factorially according to the binomial coefficient $C_{M,N} = \binom{M}{N} = \frac{M!}{N! (M-N)!}$, where $N$ is the number of nodes taken into account ($3 \leq N \leq M$).
This combinatorial increase poses computational challenges and often necessitates efficient algorithms or resourceful approximations to handle the explosive growth in complexity. Moreover, issues arise also regarding the interpretation of the vast number of high-order interactions in multiplets of different order: understanding large networks is not just about scaling up, but also about dealing with the intricate web of interactions that emerge.

\subsection{Dynamic networks of random processes: spectral analysis} \label{SpectralAnalysis}

In the linear signal processing framework, the observed network of random processes can be studied in the frequency domain in terms of the power spectral density (PSD) matrix of the stationary vector random process $Y=\{Y_1, \ldots, Y_Q\}$. The PSD matrix, denoted as $P_Y(\omega)$, is a $Q \times Q$ matrix which contains the individual PSD of the process $Y_i$,  $P_{Y_i}(\omega)$, as $i^{\mathrm{th}}$ diagonal element and the cross-PSD between the processes $Y_i$ and $Y_j$, $P_{Y_iY_j}(\omega)$, as off-diagonal elements in the position $i-j$ ($i,j=1,\ldots,Q$); $\omega \in [-\pi,\pi]$ is the normalized angular frequency ($\omega=2\pi \frac{f}{f_s}$ with $f\in [-\frac{f_s}{2},\frac{f_s}{2}]$, $f_s$ sampling frequency).
The link between the time and frequency domain representations is provided by the Fourier Transform (FT). The cross-PSD between the $i^{\mathrm{th}}$ and $j^{\mathrm{th}}$ scalar processes is defined as the FT of their cross-correlation, i.e. $P_{Y_iY_j}(\omega)=\mathfrak{F}\{R_{Y_iY_j}(k)\}$, where $R_{Y_iY_j}(k)=\mathbb{E}[Y_{i,n}Y_{j,n-k}]$; when $i=j$, the PSD of $Y_i$ is retrieved, i.e., $P_{Y_i}(\omega)=\mathfrak{F}\{R_{Y_i}(k)\}$, with $R_{Y_i}(k)=\mathbb{E}[Y_{i,n}Y_{i,n-k}]$ the autocorrelation of $Y_i$.
Furthermore, when the subdivision of $Y$ in $M$ blocks is considered, the PSD matrix $P_Y(\omega)$ can be partitioned in $M \times M$ blocks as follows:
\begin{equation}
    P_X(\omega)=\begin{bmatrix}
    P_{X_1}(\omega) & \cdots & P_{X_1 X_M}(\omega)\\
    \vdots & \ddots & \vdots\\
    P_{X_M X_1}(\omega) & \cdots & P_{X_M}(\omega)\\
    \end{bmatrix},
    \label{PSDmatrix}
\end{equation}
where the $i-j$ block has dimension $M_i \times M_j$ and is obtained as the FT of the cross-correlation matrix between $X_i$ and $X_j$, $R_{X_iX_j}(k)=\mathbb{E}[X_{i,n}^{\intercal} X_{j,n-k}]$, i.e. $P_{X_i X_j}(\omega)=\mathfrak{F}\{R_{X_iX_j}(k)\}$, with
\begin{equation}
    R_{X_iX_j}(k)= \begin{bmatrix}
    R_{X_{i_1}X_{j_1}}(k) & \cdots & R_{X_{i_1}X_{j_{M_j}}}(k)\\
    \vdots & \ddots & \vdots\\
    R_{X_{i_{M_i}}X_{j_1}}(k) & \cdots & R_{X_{i_{M_i}}X_{j_{M_j}}}(k)\\
    \end{bmatrix}.
    \label{Xcorrmatrix}
\end{equation}

The spectral densities defined above can be exploited to provide frequency domain measures of the entropy rate, mutual information rate, and O-information rate, which quantify individual, pairwise and higher-order interactions at each specific frequency.
Specifically, a spectral measure of the ER of the process $X_i$ is defined as (\cite{chicharro2011spectral})
\begin{equation}
    h_{X_i}(\omega)=\frac{1}{2}\log (2\pi e)^{M_i}|P_{X_i}(\omega)|,
    \label{spectralER}
\end{equation}
and a spectral measure of the MIR between the processes $X_i$ and $X_j$ is defined as (\cite{antonacci2021measuring})
\begin{equation}
    i_{X_i;X_j}(\omega)=\frac{1}{2}\log \frac{|P_{X_i}(\omega)| |P_{X_j}(\omega)|}{|P_{[X_iX_j]}(\omega)|},
    \label{spectralMIR}
\end{equation}
where
\begin{equation}
    P_{[X_iX_j]}(\omega)=\begin{bmatrix}
    P_{X_i}(\omega) & P_{X_i X_j}(\omega)\\
    P_{X_j X_i}(\omega) & P_{X_j}(\omega)\\
    \end{bmatrix},
    \label{PSDmatrixXiXj}
\end{equation}
and $|\cdot|$ stands for matrix determinant.
Moreover, a spectral measure of the OIR among $N$ processes $X^N=\{X_1,\ldots,X_{N}\}$ is defined in iterative form using gradients and expanding in the frequency domain the time domain expressions (\ref{OIR}) and (\ref{DeltaOIR}) as (\cite{faes2022new}):
\begin{align}
   \nu_{X^N}(\omega) &=\nu_{X^{N-1}}(\omega) + \delta_{X_N;X^{N-1}}(\omega), \label{spectralOIR} \\
   \delta_{X_N;X^{N-1}}(\omega) &=\sum_{i=1}^{N-1}i_{X_N;X^{N-1}_{-i}}(\omega)+(2-N)i_{X_N;X^{N-1}}(\omega).
   \label{spectralDeltaOIR}
\end{align}
Importantly, the spectral information measures of the interactions of order one in a process, of the pairwise interactions between two processes, and of the higher-order interactions among more than two processes, defined respectively by the spectral ER (\ref{spectralER}), by the spectral MIR (\ref{spectralMIR}) and by the spectral OIR (\ref{spectralOIR}) and OIR gradient (\ref{spectralDeltaOIR}), are closely related to the time domain information measures defined in Eqs. (\ref{ER},\ref{MIR},\ref{OIR},\ref{DeltaOIR}) by means of the so-called spectral integration property. This property shows that the integration over all frequencies of a spectral measure yields the corresponding time domain measure (\cite{geweke1982measurement, chicharro2011spectral}):
\begin{align}
    H_{X_i}=\dfrac{1}{2\pi} \int_{-\pi}^{\pi} h_{X_i}(\omega) \,\textrm{d}\omega, \label{SpectIntER} \\
    I_{X_i;X_j}=\dfrac{1}{2\pi} \int_{-\pi}^{\pi} i_{X_i;X_j}(\omega) \,\textrm{d}\omega, \label{SpectIntMIR} \\
    \Omega_{X^N}=\dfrac{1}{2\pi} \int_{-\pi}^{\pi} \nu_{X^N}(\omega) \,\textrm{d}\omega, \label{SpectIntOIR} \\
    \Delta_{X_N;X^{N-1}}=\dfrac{1}{2\pi} \int_{-\pi}^{\pi} \delta_{X_N;X^{N-1}}(\omega) \,\textrm{d}\omega. \label{SpectIntOIRgradient}
\end{align}
The spectral integration property is very important not only to connect the time and frequency domain formulations of the interaction measures at different orders, but also to allow quantification of these measures with reference to specific oscillatory components contained within spectral bands of interest. Examples of band-specific integration of the spectral interaction measures to obtain values related to peculiar oscillations of a group of random processes are reported in the next sections for both simulated and physiological network systems.

The software and the codes relevant to this work are collected in the \textit{SIR} Matlab toolbox and available for free download from \url{https://github.com/LauraSparacino/SIR_toolbox}.

\section{Practical Implementation}\label{practical_implementation}
This section presents technical tools for the practical implementation of the interaction measures presented in Sect. \ref{DynamicAnalysis} and Sect. \ref{SpectralAnalysis}.
Specifically, two methods for the estimation of the PSD of the original process $Y$ will be reviewed (Sect. \ref{estimation_methods}), i.e., non-parametric (Sect. \ref{non_parametric_estim}) and parametric (Sect. \ref{parametric_estim}) estimators. The practical computation of the ER, MIR and OIR measures in the time and frequency domain will be detailed in Sect. \ref{computation}. References to the toolbox functions performing the described calculations will be provided.
In Sect. \ref{statistical_significance}, the statistical validation of the proposed measures will be discussed.

\subsection{Estimation methods}\label{estimation_methods}
Power spectral density estimation is a crucial task in analyzing the frequency content of the investigated processes. Furthermore, as the proposed measures of information rate can be computed entirely from the elements of the PSD matrix of the original process $Y$, good PSD estimates must be provided to cope with the challenge of suitably interpreting these measures.
There are various methods to estimate PSD, each with its own strengths and weaknesses (\cite{kay1988modern, pinna1996application, zhao2019nonparametric}). The choice of the method depends on the characteristics of the signal, the available data, and the specific requirements of the analysis, being often a trade-off between frequency resolution, variance reduction, and computational complexity.
In the following sections, we discuss two well-known estimation methods, i.e., a non-parametric estimator (Sect. \ref{non_parametric_estim}) and a parametric estimator (Sect. \ref{parametric_estim}).

\subsubsection{Non-parametric estimator}\label{non_parametric_estim}
A well-known non-parametric approach to derive the PSD of the original process $Y$ is the weighted covariance (WC) method, which exploits the FT of the sample autocorrelation and cross-correlation functions of the data (\cite{blackman1958measurement}). 
By partitioning the PSD matrix of the process $Y$ in $M$ blocks as in (\ref{PSDmatrix}), and considering that the cross-correlation matrix between the two generic blocks $X_i$ and $X_j$ can be computed as in (\ref{Xcorrmatrix}), the WC estimator of the PSD computes the cross-PSD between $X_i$ and $X_j$ as
\begin{equation}
    \hat{P}_{X_i X_j}(\omega)=\sum_{k=-\tau}^{\tau}w(k)\hat{R}_{X_i X_j}(k)e^{-\mathbf{j} \omega k}, 
    \label{WC_estimator}
\end{equation}
where $\mathbf{j}=\sqrt{-1}$.
The autocorrelation of $X_i$ and the corresponding PSD are obtained when $i=j$. With $L$ being the number of data samples available, $\tau \leq L-1$ is the maximum lag for which the correlation is estimated, and $w$ is a lag window of width $2\tau$ ($w(k)=0$ for $|k|>\tau$), normalized ($0\leq w(k) \leq w(0)=1$) and symmetric ($w(-k)=w(k)$) (\cite{kay1988modern}).

Working with biased estimators for both cross-correlation and autocorrelation functions guarantees semi-definite sequences and thus does not lead to negative spectral estimates. A biased estimator of the cross-correlation function can be defined as 
\begin{equation}
    \hat{R}_{X_i X_j}(k)=\dfrac{1}{L} \sum_{n=0}^{L-1-k} X_{i,n}^{*}X_{i,n+k}, 
\end{equation}
where $^*$ stands for conjugate transpose. Remarkably, the latter holds for $k=0,\dots, L-1$; if $k=-(L-1),\dots,-1$, the auto-covariance matrix is defined as $\hat{R}_{X_i X_j}(k)=\hat{R}_{X_i X_j}^{*}(-k)$. 

Window selection is usually performed by looking at the spectral leakage introduced by the profile of the window itself (\cite{pinna1996application}). Following this rationale, the Parzen window can be suitably selected, since it shows a significantly lower side-lobe level compared to Hanning and Hamming windows; furthermore, it is non-negative for all frequencies, and produces non-negative spectral estimates (\cite{priestley1981spectral}). For the Parzen window, the relationship between the bandwidth ($B_w$) of the spectral window and the lag $\tau$ at which correlation estimates are truncated is $B_w=1.273 f_s/ \tau$. To resolve the corresponding peaks in the spectrum, the window bandwidth can be set equal to $25$ Hz, which brings to $\tau \approx 0.05$ using $f_s=1$ Hz.

\subsubsection{Parametric estimator}\label{parametric_estim}
This section reports the parametric formulation of the PSD matrix of the stationary vector random process $Y$, whose linear representation is provided by the vector autoregressive (VAR) model (\cite{lutkepohl2005new}):
\begin{equation}
   Y_n = \sum_{k=1}^{p}\mathbf{A}_k Y_{n-k} + U_n
   \label{VAR_model},
\end{equation}
where $p$ is the model order, defining the maximum lag used to quantify interactions, $Y_n=[Y_{1,n} \cdots Y_{Q,n}]^\intercal$ is a $Q$-dimensional vector collecting the present state of all processes, $\mathbf{A}_k$ is the $Q \times Q$ matrix of the model coefficients relating the present with the past of the processes at lag $k$, and $U_n=[U_{1,n} \cdots U_{Q,n}]^\intercal$ is a vector of $Q$ zero-mean white noises with $Q \times Q$ positive definite covariance matrix $\mathbf{\Sigma}_U=\mathbb{E}[U_n U_n^\intercal]$.

The VAR model (\ref{VAR_model}) provides a global representation of the overall multivariate process, and can be exploited to compute interaction measures in the frequency domain. To this end, the FT of (\ref{VAR_model}) is taken to derive
\begin{equation}
   Y(\omega) = [\mathbf{I} - \sum_{k=1}^{p}\mathbf{A}_k e^{-\mathbf{j} \omega k}]^{-1} U(\omega)=\mathbf{H}(\omega)U(\omega)
   \label{VAR_model_freq},
\end{equation}
where $Y(\omega)$ and $U(\omega)$ are the Fourier transforms of $Y_n$ and $U_n$, 
$\mathbf{j}=\sqrt{-1}$ and $\mathbf{I}$ is the $Q$-dimensional identity matrix. The $Q \times Q$ matrix $\mathbf{H}(\omega)$ contains the transfer functions relating the FTs of the innovation processes in $U$ to the FTs of the processes in $Y$.
When the subdivision of $Y$ in $M$ blocks is considered to yield $X=\{X_1,\ldots,X_M\}$, this matrix can be partitioned in $M \times M$ blocks to evidence the spectral properties related to the internal dynamics, through the $M_i \times M_i$ diagonal blocks $\mathbf{H}_{ii}(\omega)$, or to the causal interactions between $X_i$ and $X_j$, through the $M_i \times M_j$ off-diagonal blocks $\mathbf{H}_{ij}(\omega)$, $i,j\in \{1,\ldots,M\}, i \neq j$. The same partitioning is applied to the $Q \times Q$ PSD matrix of the process, defined in Sect. \ref{SpectralAnalysis} and shown in (\ref{PSDmatrix}).
Then, using spectral factorization, the PSD of $X$ can be expressed as $P_X(\omega)=\mathbf{H}(\omega)\mathbf{\Sigma}_{U}\mathbf{H}^*(\omega)$, where $^*$ stands for conjugate transpose. 

The identification procedure of the VAR model (\ref{VAR_model}) is typically performed by means of estimation methods based on minimizing the prediction error, i.e., the difference between actual and predicted data (\cite{kay1988modern, lutkepohl2005new}). While several approaches have been proposed throughout the years (\cite{schlogl2006comparison, antonacci2020information}), the most common estimator is the multivariate version of the ordinary least-squares (OLS) method (\cite{lutkepohl2005new}). Briefly, defining the past history of $Y$ truncated at $p$ lags as the $pQ$-dimensional vector $Y_n^p=[Y_{n-1}^\intercal,\ldots,Y_{n-p}^\intercal]^\intercal$ and considering $L$ consecutive time steps, a compact representation of (\ref{VAR_model}) can be defined as $\textbf{y}=\mathbf{A}\textbf{y}^p+\textbf{U}$, where $\mathbf{A}=[\mathbf{A}_1, \dots, \mathbf{A}_p]$ is the $Q \times pQ$ matrix of unknown coefficients, $\textbf{y}=[Y_{p+1},\ldots,Y_{L}]$ and $\textbf{U}=[U_{p+1},\dots,U_{L}]$ are $Q \times (L-p)$ matrices, and $\textbf{y}^p=[Y_{p+1}^p,\ldots,Y_{L}^p]$ is a $pQ\times(L-p)$ matrix collecting the regressors. 
The method estimates the coefficient matrices through the OLS formula, $\widehat{{\mathbf{A}}}=\textbf{y}(\textbf{y}^p)^\intercal[\textbf{y}^p(\textbf{y}^p)^\intercal]^{-1}$. The innovation process is estimated as the residual time-series $\widehat{\textbf{U}}=\textbf{y}-\widehat{{\mathbf{A}}}\textbf{y}^p$, whose covariance matrix $\widehat{\boldsymbol{\Sigma}}_{U}$ is an estimate of $\boldsymbol{\Sigma}_U$.
By partitioning the PSD matrix of the process $Y$ in $M$ blocks as in (\ref{PSDmatrix}), the latter can be estimated as $\widehat{P}_{X}(\omega)=\widehat{\mathbf{H}}(\omega) \widehat{\boldsymbol{\Sigma}}_{{U}}\widehat{{\mathbf{H}}}^{*}(\omega)$, where the transfer matrix is $\widehat{{\mathbf{H}}}(\omega)=[\textbf{I}-\widehat{\mathbf{A}}(\omega)]^{-1}$, with $\widehat{\mathbf{A}}(\omega) = \sum_{k=1}^{p}\widehat{\mathbf{A}}_k e^{-\mathbf{j} \omega k}$. \\
As regards the selection of the model order $p$, several criteria exist for its determination (see, e.g., \cite{lutkepohl2005new, karimi2011order}).
One commonly used approach is to set the order according to the Akaike Information Criterion (AIC) (\cite{akaike1974new}), or the Bayesian information criterion (BIC) (\cite{schwarz1978estimating}). Practically, it is crucial to find the right balance between excessively low orders, which might lead to an inadequate description of crucial oscillatory information in the vector process, and overly high orders, which could result in overfitting, with the outcome that the model captures not only the desired information but also includes noise.

\subsection{Computation of interaction measures}\label{computation}
The practical computation of time and frequency domain measures of information rate from a set of $Q$ time series of $L$ samples, $\textbf{y}_i=\{y_{i}(1),\ldots,y_{i}(L)\}$, where $i=1,\ldots,Q$ and $L$ is the length of the time series, measured from a physical system, is based on considering the series as a finite length realization of the vector process $Y = \{ Y_1, \ldots, Y_Q \}$ that describes the evolution of the system over time.
Then, computation of the spectral measures of entropy rate (\ref{spectralER}), mutual information rate (\ref{spectralMIR}), and O-information rate (\ref{spectralOIR}) amounts to estimate the power spectral density matrix of the set $\textbf{y}=\{ \textbf{y}_1, \ldots, \textbf{y}_Q \}$ through the non-parametric (Sect. \ref{non_parametric_estim}, function \texttt{sir\_WCspectra.m}) or the parametric (Sect. \ref{parametric_estim}, function \texttt{sir\_VARspectra.m}) methods. The time domain counterparts of the ER, MIR and OIR measures can be obtained exploiting (\ref{SpectIntER}), (\ref{SpectIntMIR}) and (\ref{SpectIntOIR}), respectively. The functions \texttt{sir\_mir.m} and \texttt{sir\_oir.m} compute the time and frequency domain ER and MIR measures, and OIR measures, respectively, accepting as input the PSD matrix of $\textbf{y}$. Specifically, while the non-parametric estimation method is based on computing the PSD directly from the raw data through the FT of the windowed correlation function, the linear VAR equation in (\ref{VAR_model}) is seen as a model of how the observed data have been generated, and an identification procedure (function \texttt{sir\_idMVAR.m}) is applied after model order selection (function \texttt{sir\_mos\_idMVAR.m}) to provide estimates of the coefficients and innovation variances to be used for computing the information measures.

\subsection{Assessment of Statistical Significance}\label{statistical_significance}
This section presents the use of surrogate and bootstrap data analyses to statistically validate the proposed measures of information rate. Validation is performed at the level of individual realizations of the observed variables $\{ Y_1, \ldots, Y_Q \}$, obtained in the form of the set of time series $\textbf{y}_i=\{y_{i}(1),\ldots,y_{i}(L)\}$, where $i=1,\ldots,Q$ and $L$ is the length of the time series.

\subsubsection{Surrogate data analysis}
The method of surrogate data (\cite{theiler1992}) is employed to set a significance level for the ER and MIR measures.

Statistical validation of the ER measures is performed by generating randomly shuffled surrogates (\cite{palus1997}), which are realizations of independent and identically distributed (i.i.d.) stochastic processes with the same mean, variance and probability distribution as the original series, obtained by randomly permuting in temporal order the samples of the original series, according to the null hypothesis of fully unpredictable process without temporal statistical structure (function \texttt{sir\_surrshuf.m}).
This procedure is repeated $N_s$ times to obtain the surrogate series $\textbf{y}_i^s$ ($i=1,\ldots,Q; s=1,\ldots,N_s$). The time domain ER is then estimated on each surrogate, yielding the surrogate distribution from which the significance threshold is derived taking the $100(\alpha)^{\mathrm{th}}$ percentile, where $\alpha$ is the prescribed significance level. The original time domain ER value is deemed as statistically significant if it stands below the threshold. As regards the spectral counterparts, the spectral ER profiles are estimated on each surrogate and integrated to get surrogate ER values in specific frequency bands, yielding the surrogate distributions from which the significance thresholds are derived taking the $100(\frac{\alpha}{2})^{\mathrm{th}}$ and $100(1-\frac{\alpha}{2})^{\mathrm{th}}$ percentiles. The original frequency domain ER value, i.e., the value obtained from integration of the spectral ER profile within a given frequency band, is deemed as statistically (non) significant if it stands (below) above the (lower) higher threshold.

To statistically validate MIR measures, surrogate time series which preserve the individual linear correlation properties of two series but destroy any correlation between them are obtained through the iterative Amplitude Adjusted Fourier Transform (iAAFT) procedure (\cite{schreiber1996surrogate}), which represents an advancement over the FT method (function \texttt{sir\_surriaafft.m}). It generates surrogate time series by computing the FT of the original series, substituting the Fourier phases with random numbers uniformly distributed between 0 and $2\pi$, and then performing the inverse FT. 
This procedure is repeated $N_s$ times to obtain the set of surrogate series $\textbf{y}_i^s$ ($i=1,\ldots,Q; s=1,\ldots,N_s$). The MIR is then estimated on each surrogate pair $\{ \textbf{y}_i^s, \textbf{y}_j^s \}$ ($i,j=1,\ldots,Q, i \neq j$), yielding the surrogate distribution from which the significance threshold is derived taking the $100(1-\alpha)^{\mathrm{th}}$ percentile. Finally, the original MIR value is deemed as statistically significant if it stands above the threshold. The same procedure applies to both time and frequency domain MIR values, the latter obtained by integrating the spectral MIR profiles within specific frequency bands.

\subsubsection{Bootstrap data analysis}
The bootstrap method (\cite{efron1979bootstrap_first}) is exploited to identify confidence intervals for the OIR measures and thus assess their statistical significance.
For this purpose, the block bootstrap data generation procedure (\cite{politis2003bootstrap}) is followed to generate, starting from the original time series, $N_s$ bootstrap pseudo-series $\textbf{y}_i^s$ ($i=1,\ldots,Q; s=1,\ldots,N_s$), which maintain all the features of the original time series, i.e., individual and coupling properties (function \texttt{sir\_block\_bootstrap.m}).
For further technical details about the procedure, see (\cite{sparacino2023statistical}).
Then, the OIR is computed at each order $N$ from the new, full-size bootstrap series $\textbf{y}_i^s$. The procedure is iterated $N_s$ times to construct bootstrap distributions, whose confidence intervals are exploited to check the statistical significance of the original OIR values. Specifically, when a given bootstrap distribution comprises the zero at the $\alpha$ significance level, i.e., if the zero value is below the $100(1-\frac{\alpha}{2})^{\mathrm{th}}$ and above the $100(\frac{\alpha}{2})^{\mathrm{th}}$ percentile of that distribution, the corresponding OIR measure is deemed as not statistically significant. The same procedures applies to both time and frequency domain OIR values, the latter obtained by integrating the spectral OIR profiles within specific frequency bands.

\section{Theoretical Examples}\label{th_simulations}
This section reports illustrative examples of simulated VAR processes, used as a benchmark to illustrate the properties of the time and frequency domain information-theoretic measures described in Sect. \ref{sec_methods} and here computed from the true values imposed for the VAR parameters.
Specifically, we consider $Q$ linearly interacting Gaussian processes $Y=\{ Y_1, \ldots, Y_Q\}$ grouped in $M$ blocks, where the process $Y_i$, $i=1,\ldots,Q$, is defined as: 
\begin{equation}
Y_{i,n} = \sum_{k=1}^{p} \sum_{j=1}^Q a_{ij,k}Y_{j,n-k} + U_{i,n},
\label{sim_VAR_process}
\end{equation}
with $p$ the maximum model order used to describe lagged interactions, $a_{ij,k}$ the coefficient describing the causal interaction from $Y_{j,n-k}$ to $Y_{i,n}$ at lag $k$, and $U_i$ a Gaussian white noise process with zero mean and unit variance. The autonomous oscillations of $Y_i$ are obtained placing pairs of complex-conjugate poles, with modulus $\rho$ and phase $2\pi{f}$, in the complex plane representation of the process; the AR coefficients resulting from this setting are $a_{ii,1}=2\rho \cos(2\pi f)$ and $a_{ii,2}=-\rho^2$ (\cite{faes2015information}). The whole VAR process $Y$ can be defined in compact form as in (\ref{VAR_model}). \\
Starting from this representation, herein we analyse two different simulation settings.
In Sect. \ref{sim1}, we study the behaviour of a \textit{star} structure centred on the process $Y_1$, where incoming or outgoing causal interactions from/to the remaining processes are selectively turned on to reproduce the prevalence of synergistic or redundant characters of interplay in the analyzed network. Conversely, in Sect. \ref{sim2}, we show how redundancy and synergy can coexist at different orders in the spectral domain, and how this harmony may remain hidden if the whole-band integration of the proposed frequency-specific measures is performed. 

\subsection{Gaussian processes interacting in star structures}\label{sim1}
\begin{figure*}
    \centering \includegraphics{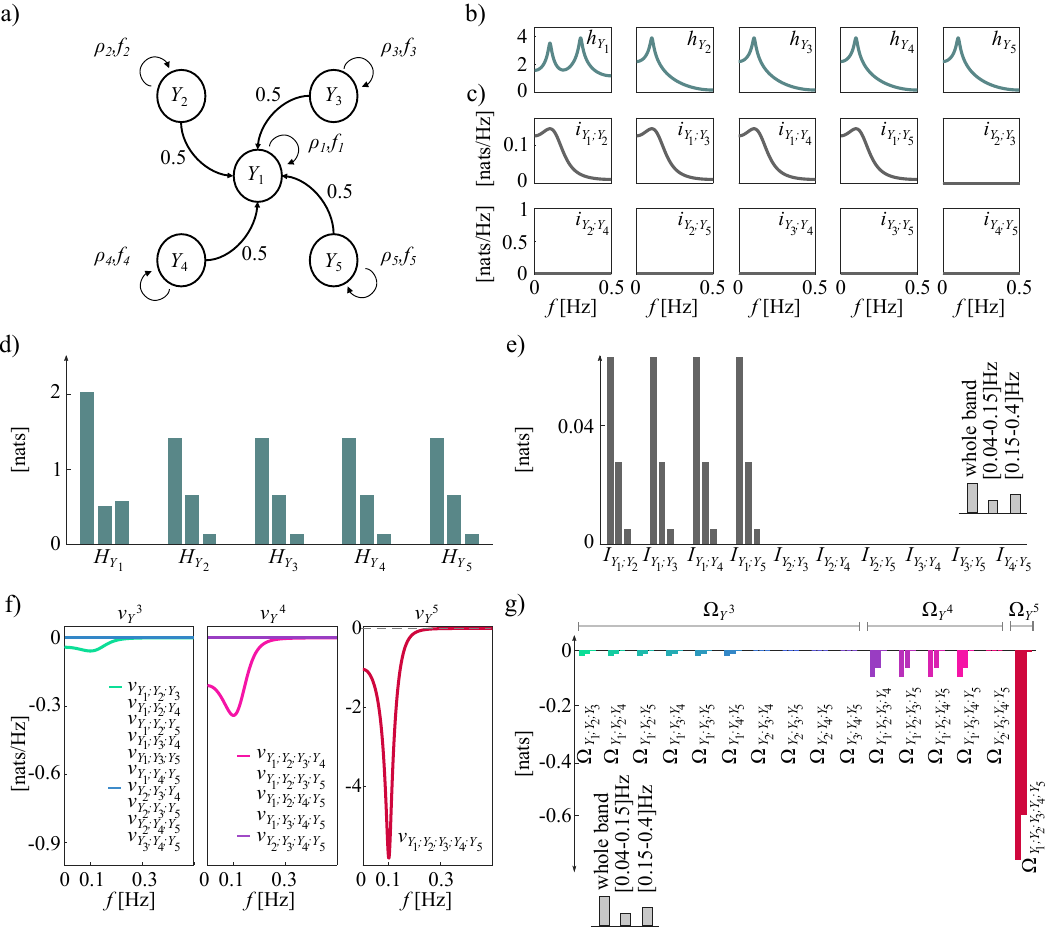}
    \caption{ \textbf{Synergy arises when multiple processes send unique information to the same target}. \textbf{a)} Simulation design, where $\rho_i$ is the radius and $f_i$ is the oscillating frequency of the process $Y_i$ ($i=1,\ldots,5$). \textbf{b)} Spectral entropy rates $h_{Y_i}$ of the processes $Y_i$ ($i=1,\ldots,5$). \textbf{c)} Spectral mutual information rates $i_{Y_i;Y_j}$ between the processes $Y_i$ and $Y_j$ ($i,j=1,\ldots,5$, $i \neq j$). \textbf{d)} Entropy rate values integrated in the whole band (left bars), the range $[0.04 - 0.15]$ Hz (middle bars) and the range $[0.15 - 0.4]$ Hz (right bars) of the spectrum. \textbf{e)} Mutual information rate values integrated in the whole band (left bars), the range $[0.04-0.15]$ Hz (middle bars) and the range $[0.15-0.4]$ Hz (right bars) of the spectrum. 
    \textbf{f)} Spectral O-information rates of order 3 ($\nu_{Y^3}$, left), 4 ($\nu_{Y^4}$, middle), and 5 ($\nu_{Y^5}$, right). \textbf{g)} O-information rate values of order 3 ($\Omega_{Y^3}$), 4 ($\Omega_{Y^4}$), and 5 ($\Omega_{Y^5}$) integrated in the whole band (left bars), the range $[0.04-0.15]$ Hz (middle bars) and the range $[0.15-0.4]$ Hz (right bars) of the spectrum. 
    }
    \label{fig:th_sim_star_in}
\end{figure*}

\begin{figure*}
    \centering \includegraphics{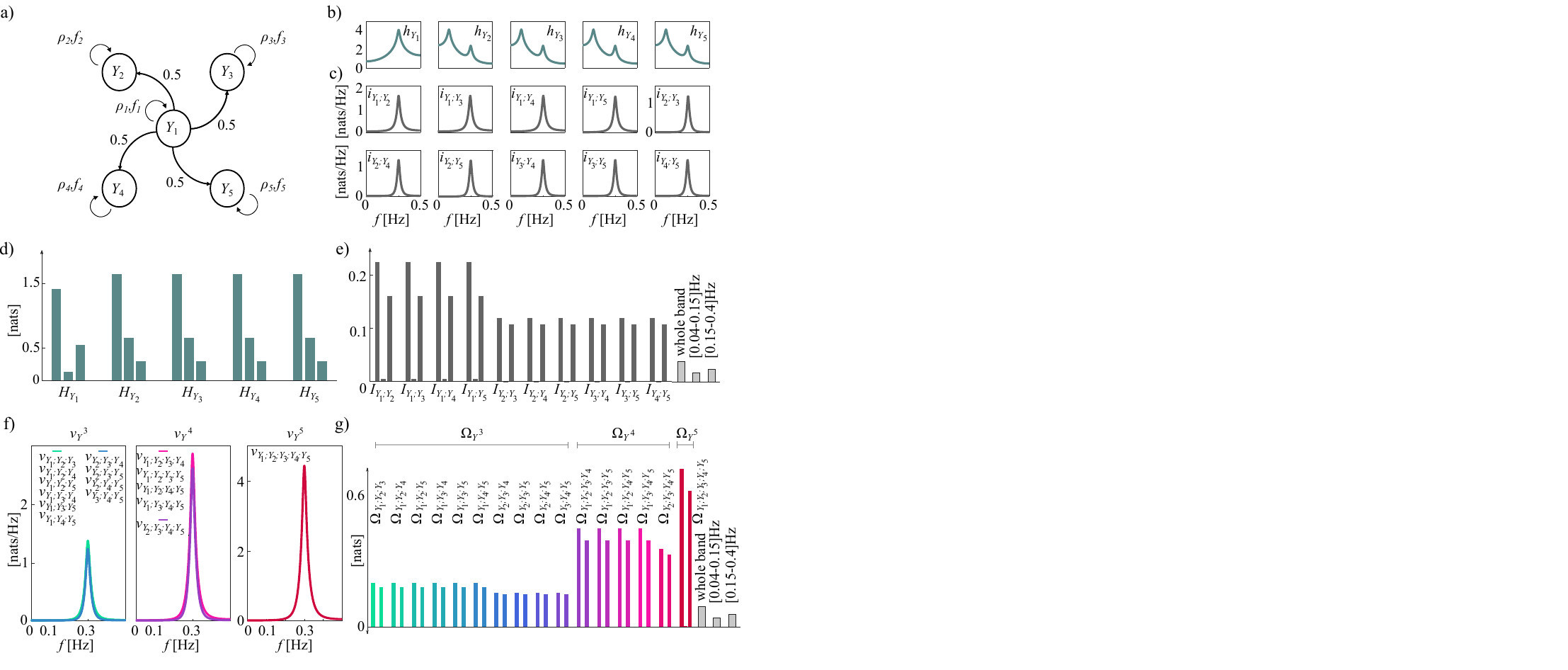}
    \caption{ \textbf{Redundancy emerges when one source process sends copies of the same information to multiple targets}. \textbf{a)} Simulation design, where $\rho_i$ is the radius and $f_i$ is the oscillating frequency of the process $Y_i$ ($i=1,\ldots,5$). \textbf{b)} Spectral entropy rates $h_{Y_i}$ of the processes $Y_i$ ($i=1,\ldots,5$). \textbf{c)} Spectral mutual information rates $i_{Y_i;Y_j}$ between the processes $Y_i$ and $Y_j$ ($i,j=1,\ldots,5$, $i \neq j$). \textbf{d)} Entropy rate values integrated in the whole band (left bars), the range $[0.04-0.15]$ Hz (middle bars) and the range $[0.15-0.4]$ Hz (right bars) of the spectrum.  \textbf{e)} Mutual information rate values integrated in the whole band (left bars), the range $[0.04-0.15]$ Hz (middle bars) and the range $[0.15-0.4]$ Hz (right bars) of the spectrum.  
    \textbf{f)} Spectral O-information rates of order 3 ($\nu_{Y^3}$, left), 4 ($\nu_{Y^4}$, middle), and 5 ($\nu_{Y^5}$, right). \textbf{g)} O-information rate values of order 3 ($\Omega_{Y^3}$), 4 ($\Omega_{Y^4}$), and 5 ($\Omega_{Y^5}$) integrated in the whole band (left bars), the range $[0.04-0.15]$ Hz (middle bars) and the range $[0.15-0.4]$ Hz (right bars) of the spectrum. }
    \label{fig:th_sim_star_out}
\end{figure*}

First, we consider $Q = 5$ stationary Gaussian stochastic processes, each describing the dynamic activity of the network nodes $\mathcal{X}_i, i=1,\ldots,M$, with $M=Q$. Their lagged interactions are mapped by a VAR process of order $p=5$, with $k=2$.
Here, we imposed autonomous oscillations for the process $Y_1$, setting $\rho_{1}=0.95, f_{1}=0.3$ Hz, so that the dynamics of $Y_1$ are determined by the fixed coefficients $a_{11,1}= -0.587, a_{11,2}= –0.9$. Conversely, the remaining processes exhibit autonomous oscillations imposed setting $\rho_{i}=0.95, f_{i}=0.1$ Hz, so that the dynamics of $Y_i$, $i=2,\ldots,Q$, are determined by the fixed coefficients $a_{ii,1}= 1.5371, a_{ii,2}= –0.9$. The sampling frequency was fixed to $f_s=1$ Hz. 
To analyze network interactions, we considered a \textit{star} structure in two different configurations, where the process $Y_1$ is (i) a \textit{receiver} (Fig. \ref{fig:th_sim_star_in}a) or (ii) a \textit{sender} (Fig. \ref{fig:th_sim_star_out}a) of information for the remaining processes $Y_i$, $i=2,\ldots,Q$. The model parameters imposed for these settings are all zero except for (i) $a_{1j,1}=0.5$, $j=2,\ldots,Q$, and (ii) $a_{i1,1}=0.5$, $i=2,\ldots,Q$, respectively. 

Results of the two simulation settings are shown in Fig. \ref{fig:th_sim_star_in}b-g and Fig. \ref{fig:th_sim_star_out}b-g, respectively. The spectral profiles of entropy rates, mutual information rates and O-information rates are shown in panels \textit{b}, \textit{c}, and \textit{f}, respectively, while in panels \textit{d}, \textit{e}, and \textit{g} the ER, MIR and OIR values integrated along the whole band (left bars), the range $[0.04-0.15]$ Hz (middle bars) and range $[0.15-0.4]$ Hz (right bars) are depicted.\\
The first configuration, shown in Fig. \ref{fig:th_sim_star_in}a, is predominantly synergistic since each of the source processes $Y_i, i=2,\ldots,Q$ sends unique information to the target process $Y_1$ at the same frequency ($0.1$ Hz). In fact, as shown in panel \textit{b}, the spectral entropy rate of the latter, i.e., $h_{Y_1}$, shows peaks not only at $0.3$ Hz, which represents its own oscillating frequency, but also at $0.1$ Hz, due to dynamic information transferred from the rest of the system. Conversely, the spectral profiles of the entropy rate of the sources, i.e., $h_{Y_i}$, are characterized by peaks at $0.1$ Hz, showing that the information content of these processes is located around that frequency. Remarkably, these findings suggest that the spectral entropy rates are characterized by frequency-specific peaks wherever the series is more predictable and owns autonomous dynamics.
These results are confirmed by the spectral behaviour of the MIRs shared between $Y_1$ and $Y_i, i=2,\ldots,Q$ ($i_{Y_1;Y_i}$), which show peaks at $0.1$ Hz, as well as between pairs of sources ($i_{Y_{s1};Y_{s2}}$; $s_1=2,\ldots,Q-1$, and $s_2=3,\ldots,Q$, with $s_1 \neq s_2$), which are null at each frequency given the absence of coupled interactions between them (Fig. \ref{fig:th_sim_star_in}c). 
The same trend is visible by looking at the spectral profiles of the $3^{rd}$-order OIR ($\nu_{Y^3}$, Fig. \ref{fig:th_sim_star_in}f), which display negative peaks at $0.1$ Hz when the analyzed multiplet includes the target $Y_1$ and two of the sources, while are null at each frequency when it includes only three isolated source processes. Correspondingly, adding one source process to multiplets already including the target increases the amount of synergy in the system, as shown by the spectral profiles of the $4^{\mathrm{th}}$- and $5^{\mathrm{th}}$-order OIR ($\nu_{Y^N}$, $N=4,5$, Fig. \ref{fig:th_sim_star_in}f).
The synergistic behaviour of the network, mainly confined to the band with central frequency $0.1$ Hz, is confirmed by the integration of the spectral profiles along given frequency ranges, showing that all the information shared between the multiple interacting processes in the network is located around $0.1$ Hz (Fig. \ref{fig:th_sim_star_in}g).\\
The second configuration, shown in Fig. \ref{fig:th_sim_star_out}a, is predominantly redundant since each of the processes $Y_i, i=2,\ldots,Q$, receives the same information from the source process $Y_1$ at $0.3$ Hz. The spectral entropy rates of the targets show peaks not only at $0.1$ Hz, which represents their own oscillating frequency, but also at $0.3$ Hz, due to dynamic information transferred from the source $Y_1$ ($h_{Y_i}$, panel \textit{b}).
This is in agreement with the spectral behaviour of the MIRs shared between pairs of processes ($i_{Y_{s1};Y_{s2}}$; $s_1=1,\ldots,Q-1$, and $s_2=2,\ldots,Q$, with $s_1 \neq s_2$, panel \textit{c}), which show peaks around $0.3$ Hz, thus confirming the frequency-specific redundant character of the multiple interactions in the analyzed network. The high-order description of the VAR process $Y$ confirms these findings, as highlighted by the positive peaks of the spectral OIR $\nu_{Y^N}$, $N=3,\ldots,5$ (Fig. \ref{fig:th_sim_star_out}f), located around $0.3$ Hz, whose amplitude increases with the size of the multiplet. Their whole-band and band-specific integration returns positive values indicating an overall prevalence of redundancy (Fig. \ref{fig:th_sim_star_out}g).

\subsection{Multiple interacting Gaussian processes analyzed in blocks}\label{sim2}
\begin{figure*}
    \centering \includegraphics{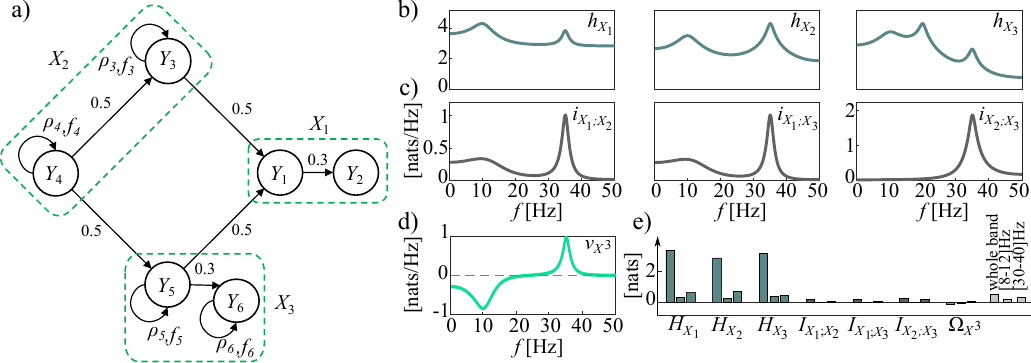}
    \caption{\textbf{Coexistence of synergy and redundancy in complex networks of interacting processes}. \textbf{a)} Simulation design, where $\rho_i$ is the radius and $f_i$ is the oscillating frequency of the process $Y_i$ ($i=1,\ldots,6$). Processes are grouped in 3 blocks $X_1, X_2, X_3$. \textbf{b)} Spectral entropy rates $h_{X_m}$ of the blocks $X_m$ ($m=1,\ldots,3$). \textbf{c)} Spectral mutual information rates $i_{X_{m_1};X_{m_2}}$ between the blocks $X_{m_1}$ and $X_{m_2}$ ($m_1,m_2=1,\ldots,3$, $m_1 \neq m_2$). \textbf{d)} Spectral O-information rate of order 3 ($\nu_{X^3}$). \textbf{e)} ER, MIR and OIR values integrated in the whole band (left bar), the range $[8 - 12]$ Hz (middle bar) and the range $[30 - 40]$ Hz (right bar) of the spectrum.}
    \label{fig:th_sim_vec}
\end{figure*}

In this simulation example, we consider $Q=6$ stationary Gaussian stochastic processes, $Y=\{ Y_1, \ldots, Y_Q\}$, grouped in $M=3$ blocks, $X=\{ X_1, \ldots, X_M\}$. Network interactions are mapped by a six-variate VAR process of order 3 configured to reproduce coexisting redundant and synergistic interactions \cite{faes2017multiscale, antonacci2020information, antonacci2021measuring}. Following the structure of the AR model in (\ref{sim_VAR_process}), autonomous oscillations in the processes $Y_i$, $i=3,\ldots,6$, are obtained placing complex-conjugate poles with radii $\rho_3=\rho_5=0.85$, $\rho_4=\rho_6=0.95$ and normalized frequencies $f_3/f_s=0.1$, $f_5/f_s=0.1$, $f_4/f_s=0.35$ and $f_6/f_s=0.2$ in the complex plane. Assuming a sampling frequency $f_s=100$ Hz, the poles determine oscillations at $10$ Hz, $35$ Hz and $20$ Hz. 
We set $X_1=\{Y_1,Y_2\}$, $X_2=\{Y_3, Y_4\}$ and $X_3=\{Y_5, Y_6\}$, and the causal interactions between different blocks are specified to obtain the common driver effect $Y_5 \leftarrow Y_4 \rightarrow Y_3$, the common child effect $Y_3 \rightarrow Y_1 \leftarrow Y_5$ and the unidirectional couplings $Y_5 \rightarrow Y_6$, $Y_1 \rightarrow Y_2$. The model parameters imposed for these settings are all zero except for $a_{15,1}=a_{54,1}=a_{13,2}=a_{34,3}=0.5$ and $a_{21,2}=a_{65,2}=0.3$, as depicted in Fig. \ref{fig:th_sim_vec}a. 

Results of the simulation are reported in Fig. \ref{fig:th_sim_vec}.
The spectral profiles of entropy rates, mutual information rates and O-information rate are shown in panels \textit{b}, \textit{c}, and \textit{d}, respectively, while in panel \textit{e} the ER, MIR and OIR values integrated along the whole band (left bars), the range $[8-12]$ Hz (middle bars) and the range $[30-40]$ Hz (right bars) are depicted.\\
In panel \textit{b}, the spectral ER profiles of the three blocks of processes are shown ($h_{X_1}$,$h_{X_2}$,$h_{X_3}$). Two distinct oscillations at $10$ Hz and $35$ Hz are consistently observed, indicating that the information content of the system is predominantly localized in specific bands of the spectrum. Specifically, $X_1$ exhibits the highest spectral information content at $10$ Hz, originating directly from the causal links $X_2\rightarrow X_1$ and $X_3\rightarrow X_1$. Conversely, the spectral ER profile of $X_2$ displays the highest peak at $35$ Hz, directly linked to the amplitude of the oscillatory activity of $Y_4$, which, in turn, is controlled by the value of $\rho_4$. Lastly, $X_3$ features three different oscillations at $10$ Hz, $20$ Hz and $35$ Hz, with the latter exhibiting the lowest amplitude since it is transferred from $X_2$ and it is not representative of the autonomous oscillatory activity of $X_3$. \\
The analysis of the spectral MIRs, whose profiles are reported in panel \textit{c}, reveals the presence of a dynamical coupling occurring at $10$ and $35$ Hz. In particular, the common drive role of $Y_4$, directed to $Y_5$ and $Y_3$, ensures the presence of a prominent peak at $35$ Hz in all the spectral MIR profiles. On the other hand, the presence of dynamic coupling at $10$ Hz between $X_1$ and $X_2$ ($i_{X_1;X_2}$), as well as between $X_1$ and $X_3$ ($i_{X_1;X_3}$), is driven by the presence of an oscillation at $10$ Hz in both $Y_3$ and $Y_5$ directly transmitted towards $Y_1$. This is not the case when analysing the MIR between $X_2$ and $X_3$, which shows only the presence of an oscillation at $35$ Hz as a result of the interaction between $Y_4$ and $Y_5$. \\
The complexity of the interactions in the analyzed network is well explained by the analysis of the spectral OIR of order 3 ($\nu_X^3$, Fig. \ref{fig:th_sim_vec}d), which reveals the coexistence of redundancy occurring at $35$ Hz due to the cascade mechanism $X_2\rightarrow X_3 \rightarrow X_1$, and synergy occurring at $10$ Hz due to the common child structure $X_2 \rightarrow X_1 \leftarrow X_3$. This is evident only when analysing the spectral profile of the OIR, but it is not detectable employing the time domain measure. Indeed, the analysis of the OIR in the time domain evidences the impossibility of discriminating the coexistence of synergistic and redundant contributions. The integrated value over the whole frequency spectrum is negative, indicating synergy; redundancy emerges only when integrating within the range $[30-40]$ Hz (Fig. \ref{fig:th_sim_vec}e).

\section{Exemplary applications to physiological networks}\label{applications}

\subsection{Cardiovascular, cardiorespiratory and cerebrovascular interactions}\label{appl_physiology}

In this section, we analyze physiological time series collected to study the effect of orthostasis on cardiovascular, cerebrovascular and respiratory variability (\cite{faes2013investigating, bari2016}). One representative subject was selected for the following analyses, chosen from a dataset comprising healthy controls enrolled at the Neurology Division of Sacro Cuore Hospital, Negrar, Italy.
Electrocardiogram was acquired together with arterial pressure (AP) measured at the level of middle finger through a photopletysmographic device (Finapres Medical Systems, Ohmenda, The Netherlands); cerebral blood flow velocity (CBFV) and respiration were measured at the level of the middle cerebral artery by means of a transcranial Doppler ultrasonographic device (Multi-Dop T2, Dwl, San Juan Capistrano, CA) and through a thoracic impedance belt, respectively. Signals were synchronously acquired at a sampling rate of 1 kHz.
From the raw signals, stationary time series of heart period (\textit{T}), systolic AP (\textit{S}), mean AP (\textit{M}), mean CBFV (\textit{F}) and respiration (\textit{R}) were measured as in (\cite{faes2013investigating, bari2016}) during the supine resting state condition, and regarded as realizations of the \textit{T}, \textit{S}, \textit{M}, \textit{F} and \textit{R} discrete-time processes, in turn assumed as uniformly sampled with a sampling frequency equal to the inverse of the mean heart period. The series, each of length equal to $250$ beats, were preprocessed reducing the slow trends with an AR high-pass filter (zero phase; cut-off frequency $0.0156$ Hz) and removing the mean value. \\
Time domain and spectral measures of ER, MIR and OIR were obtained from the non-parametric estimation of the power spectral density matrix, selecting a Parzen window with bandwidth $B_{w}=20$ Hz as in Sect. \ref{non_parametric_estim}, and then applying the derivations presented in Sect. \ref{DynamicAnalysis} and \ref{SpectralAnalysis}, as it has been detailed in Sect. \ref{computation}.
The spectral ER, MIR and OIR profiles were integrated within two frequency bands of physiological interest, i.e. the low frequency (LF, $f\in [0.04 - 0.15]$ Hz) and the high frequency (HF, $f\in [0.15 - 0.4]$ Hz) band, as well as over the whole frequency range ($f \in [0 - f_s/2]$ Hz) to get the corresponding time domain values.
Surrogate and bootstrap data analyses were applied as in Sect. \ref{statistical_significance} to assess the statistical significance of the computed measures, with $N_s=100$ iterations and $\alpha=0.05$ significance level.

\begin{figure*}
    \centering
    \includegraphics[scale=0.97]{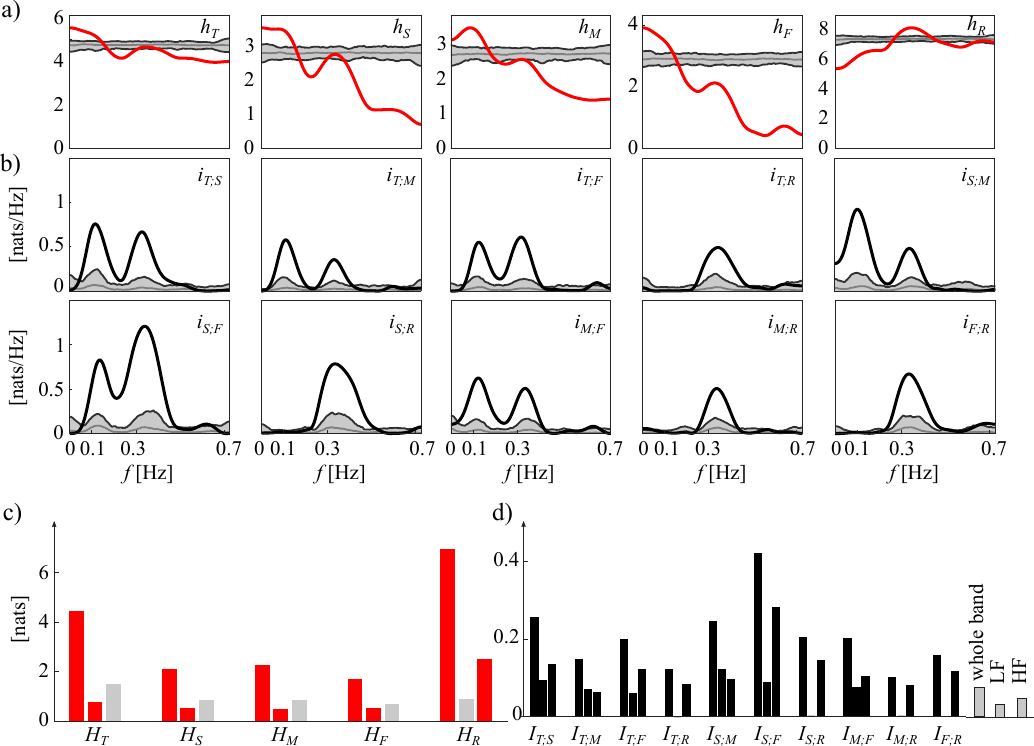}
    \caption{\textbf{Cardiovascular, cerebrovascular and respiratory signals show coherent oscillations in spectral bands with physiological meaning}.
    \textbf{a)} Red solid lines: spectral ER profiles of \textit{T} ($h_T$), \textit{S} ($h_S$), \textit{M} ($h_M$), \textit{F} ($h_F$), and \textit{R} ($h_R$) time series. \textbf{b)} Black solid lines: spectral MIR profiles for pairs of processes. The surrogate distributions of the spectral ER and MIR profiles are depicted as shaded grey areas, median (grey solid lines) and percentiles (black solid lines, computed with $5\%$ significance level).
    \textbf{c)} Entropy rate values integrated in the whole band (left bars), the low frequency (LF) band (middle bars) and the high frequency (HF) band (right bars) of the spectrum. Grey bars indicate non significant values according to surrogate data analysis. \textbf{d)} Mutual information rate values integrated in the whole band (left bars), the low frequency (LF) band (middle bars) and the high frequency (HF) band (right bars) of the spectrum. All the analyzed coupled interactions show a HF contribution due to the role of common drive exerted by respiration, while interactions directly involving respiratory variability are always non significant in the LF band of the spectrum.}
    \label{fig:appl2_er_mir}
\end{figure*}

\begin{figure*}
    \centering
    \includegraphics[scale=0.97]{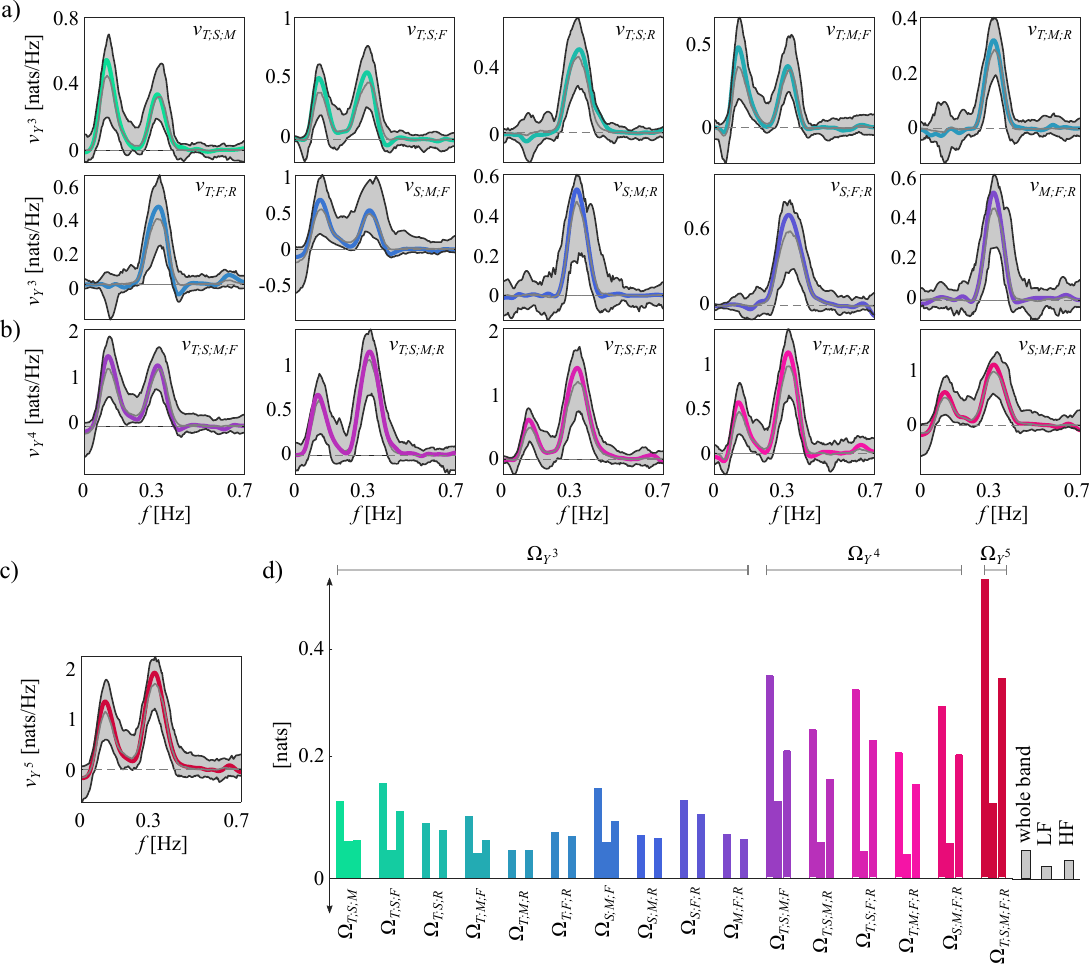}
    \caption{\textbf{Physiological interaction patterns are characterized by predominance of redundancy}.
    \textbf{a)} Spectral OIR profiles of order 3 ($\nu_{Y^3}$). \textbf{b)} Spectral OIR profiles of order 4 ($\nu_{Y^4}$). \textbf{c)} Spectral OIR profiles of order 5 ($\nu_{Y^5}$). The boostrap distributions of the spectral OIR profiles are depicted as shaded grey areas, median (grey solid lines) and percentiles (black solid lines, computed with $5\%$ significance level).
    \textbf{d)} OIR values of order 3 ($\Omega_{Y^3}$), 4 ($\Omega_{Y^4}$), and 5 ($\Omega_{Y^5}$) integrated in the whole band (left bars), the low frequency (LF, middle bars) and the high frequency (HF, right bars) band of the spectrum.
    All the analyzed high-order interactions show a HF contribution due to the role of common drive exerted by respiration, while interactions of order 3 directly involving respiratory variability are always non significant in the LF band of the spectrum. Redundancy scales with the size of the analyzed multiplet.}
    \label{fig:appl2_oir}
\end{figure*}

Results of the analysis are displayed in Fig. \ref{fig:appl2_er_mir} and Fig. \ref{fig:appl2_oir}. Panels in Fig. \ref{fig:appl2_er_mir}a show how the spectral content of the five physiologic variables, indicated by the ER profiles in red, is distributed along the whole frequency axis with respect to the corresponding thresholds, i.e., the shaded grey areas within the $2.5 \%$ (bottom black solid line) and $97.5 \%$ (top black solid line) percentiles of the ER surrogate distributions. Since the frequency-specific measures, i.e., the measures integrated along a given spectral range, are deemed as statistically (non) significant if they are (below) above the (lower) higher threshold, here we detected significant LF oscillations for \textit{T}, \textit{S}, \textit{M} and \textit{F}, as well as significant HF oscillations for the \textit{R} process only (Fig. \ref{fig:appl2_er_mir}c, grey bars indicate non significant values). These findings are expected from previous studies (\cite{pagani1986power, montano1994power, cooley1998evidence, stefanovska2002cardiorespiratory}), and confirm the presence of LF oscillatory rhythms in the variability of heart rate and blood pressure, as well as the main role exerted by the HF spectral components of the respiration signal, during the supine rest.\\
The spectral MIR profiles are shown in Fig. \ref{fig:appl2_er_mir}b. Findings suggest that respiration acts as a common driver for the other variables of the network, as indicated by the presence of significant HF peaks in all the spectral MIR profiles (black solid lines). Moreover, the absence of significant LF peaks for some of the couplings involving the \textit{R} process, i.e., $i_{T;R}$, $i_{S;R}$, and $i_{F;R}$, seems to advise that respiration leads these coupled interactions, as well as that the directed coupling $R \rightarrow Y$, where $Y=T,S,F$ may predominate over the opposite direction $Y \rightarrow R$. Importantly, the low-frequency descriptors of interactions involving only the processes $T$, $S$, $M$, and $F$ are all statistically significant, highlighting the sympathetic nature of the oscillations of these physiologic variables (\cite{pagani1986power, montano1994power, zhang2000spontaneous}).\\
Panels in Fig. \ref{fig:appl2_oir}a-c display the spectral profiles of the OIR of order 3 ($\nu_{Y^3}$, panel \textit{a}), 4 ($\nu_{Y^4}$, panel \textit{b}) and 5 ($\nu_{Y^5}$, panel \textit{c}). Previous results are confirmed here, i.e., triplets including the \textit{R} process display non-significant LF oscillations but significantly redundant HF oscillations, supporting the hypothesis of the role of common drive exerted by respiration (\cite{bari2016nonlinear, faes2022quantifying, sparacino2022quantifying, scagliarini3gradients}). This trend is confirmed in the case of multiplets of order 4.
Remarkably, the major contribution to the time domain OIR values is brought by the HF band independently of the considered multiplet, as evidenced by bars in panel \textit{d}. Furthermore, it is worth noting that redundancy scales with the size of the multiplet, again corroborating the hypothesis of prevalence of common drive effects led by respiration variability within the analyzed physiological network.

\subsection{Brain Dynamics}\label{appl_brain}

In this section, we analyze electroencephalographic (EEG) signals relevant to one healthy subject performing a motor execution task; data is available at \href{https://physionet.org/content/eegmmidb/1.0.0/}{https://physionet.org/content/eegmmidb}. The dataset comprises 64 EEG electrodes referenced to both mastoids (international 10-20 system, $f_s=160$ Hz) (\cite{schalk2004bci2000, goldberger2000physiobank}). The subject was asked to open and close the right fist cyclically until a target on the right side of a screen disappeared. The raw signals were firstly detrended, then filtered (band-pass, 2-45 Hz; notch, 59-61 Hz) and finally epoched to extract 20 trials of 4 s each. All trials were then reduced to zero mean and unit variance. We selected 6 EEG electrodes located over the contralateral and ipsilateral motor areas, and grouped them in 3 blocks, i.e. $X_1=[C_3,C_1]$, $X_2=[C_2,C_4]$ and $X_3=[F_z,FC_z]$. For each trial, a VAR model was identified through vector least-squares, fixing the model order to 10. Then, the estimated VAR parameters were used to compute the spectral ER, MIR and OIR profiles, as detailed in Sect. \ref{SpectralAnalysis} and discussed for practical implementations in Sect \ref{computation}. Finally, time domain counterparts for the $\alpha$ and $\beta$ brain rhythms, as well as over the whole frequency range were obtained by integrating the interaction measures over the relevant frequency ranges (i.e., $\alpha=[7-15]$ Hz, $\beta=[18-26]$ Hz, $f \in [0-f_s/2]$ Hz, respectively). Surrogate and bootstrap data analyses were applied as in Sect. \ref{statistical_significance} to assess the statistical significance of the computed measures, with $N_s=100$ iterations and $\alpha=0.05$ significance level.

\begin{figure*}
   \centering
  \includegraphics{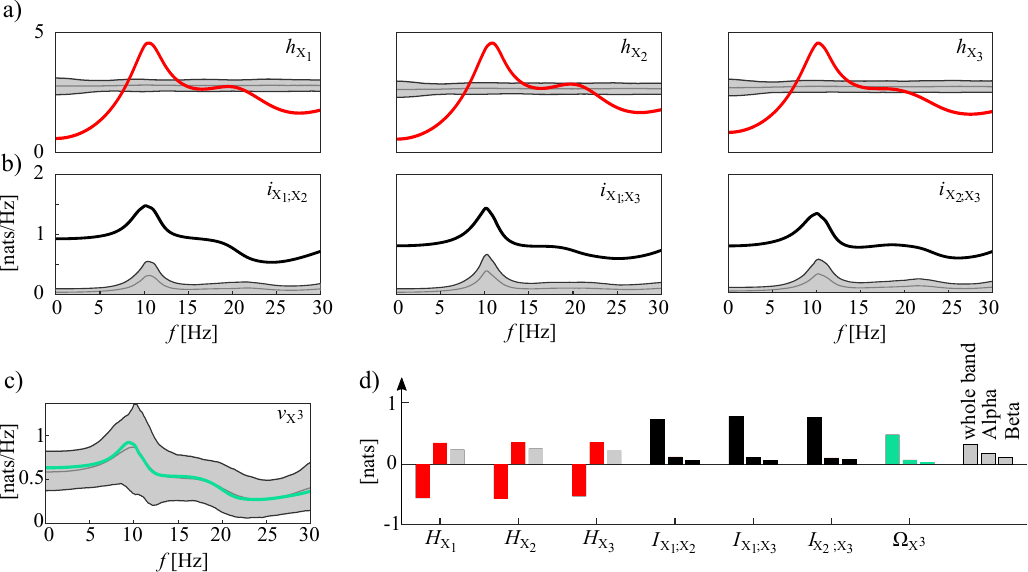}
   \caption{\textbf{Brain dynamics during motor imagery are characterized by predominance of redundancy}.
    \textbf{a)} Spectral ER profiles ($h_{X_1}$,$h_{X_2}$,$h_{X_3}$). \textbf{b)} Spectral MIR profiles ($i_{X_1;X_2}$,$i_{X_1;X_3}$,$i_{X_2;X_3}$). The surrogate distributions of the spectral ER and MIR profiles are depicted as shaded grey areas, median (grey solid lines) and percentiles (black solid lines, computed with $5\%$ significance level).
    \textbf{c)} Spectral OIR of order 3 ($\nu_{X^3}$). The boostrap distribution is depicted as shaded grey area, median (grey solid line) and percentiles (black solid lines, computed with $5\%$ significance level).
    \textbf{d)} ER, MIR and OIR values integrated in the whole band (left bars), the alpha ($\alpha$, middle bars) and the beta ($\beta$, right bars) frequency bands of the spectrum.}
   \label{fig:eeg_oir}
\end{figure*}

Fig. \ref{fig:eeg_oir} reports the grand-average over trials of the frequency profiles of ER (panel \textit{a}), MIR (panel \textit{b}), and OIR (panel \textit{c}), computed during the motor execution task and depicted over the frequency range $f \in [1-30]$ Hz. The spectral ERs (panel \textit{a}), compared with the $97.5^{\mathrm{th}}$ percentile of the surrogate distributions (top black solid lines), show statistically significant oscillations around 10 Hz ($\alpha$ band). The oscillations in $\beta$ band were found to be non significant since they are all below the $2.5^{\mathrm{th}}$ percentile of the surrogate distributions. Integrated values of ER demonstrate the presence of oscillations in $\alpha$ and $\beta$ bands as well (panel \textit{d}), in accordance with the physiology of the motor execution as demonstrated in several works on this topic (\cite{cona2009changes, antonacci2021measuring, pirovano2022resting}).\\
The spectral MIRs and their integrated values are shown in panels \textit{b} and \textit{d}, respectively. The MIR shows statistical significance across all potential pairs of processes within both $\alpha$ and $\beta$ bands. Although the difference is not so apparent when compared to other possible pairs, the highest MIR value emerges when examining the interaction between $X_1$ and $X_2$ ($i_{X_1;X_2}$). This underscores the presence of a robust dynamical coupling between the two brain hemispheres during the execution of a motor task (\cite{grefkes2008dynamic}).\\
The spectral OIR and the corresponding integrated values in the time domain, $\alpha$ and $\beta$ bands are reported in panels \textit{c} and \textit{d}. The study of the interaction of order 3 in both time and frequency domain reveals a statistically significant redundant contribution in the network with a prominent peak in the $\alpha$ band. This result can be related with the prevalence of redundancy in EEG dynamics during motor task execution, previously highlighted in other studies (\cite{antonacci2021measuring, pirovano2023rehabilitation}). Remarkably, the dominance of redundancy may be ascribed to the effects of volume conduction that blur the information identified at the level of scalp EEG sensors (\cite{van2019critical}).

\section*{Conclusions}
This work unifies several information-theoretic approaches for the study of node-specific, pairwise and high-order dynamical interactions in network systems. Flexibility and scalability of the proposed framework are guaranteed by the utilization of information-theoretic measures defined for scalar or vector processes, in both time and frequency domains in a way such that the two representations are tightly connected in a straightforward way, i.e., by satisfaction of the spectral integration property.
Network interactions are categorized hierarchically depending on the number of nodes involved in the computation of each interaction measure: entropy rate describes the predictable information within a node, mutual information rate describes the information dynamically shared between two nodes, interaction information and O-information rates describe the information shared among three or more nodes through concepts of redundancy and synergy, thus opening the way to a deeper investigation of hHONs.
Crucially, the expansion of each measure in the frequency domain allows to provide a spectral representation of the information processed in the analyzed network, as well as to focus on specific frequency bands related to oscillations with physiological meaning.

The application of the framework to physiological time series illustrates its capability to capture the complexity of the dynamics at the level of nodes, as well as the tangle of interactions among pairs and group of nodes. Remarkably, the utilization of measures of high-order interactions allows to describe and quantify the balance between redundancies and synergies among arbitrarily large groups of nodes of physiological networks. Moreover, our work highlights the importance of studying these features within specific frequency bands of biological interest to elicit interactions which may be otherwise hidden if investigated only in the time domain.
The unified design of the framework where measures of interaction at different orders are formulated uniformly in the time and frequency domains, together with the availability of estimation algorithms and related codes, will foster the diffusion of these measures in a range of applications in the fields of Network Neuroscience and Network Physiology. 

\bibliography{SIRarXiV}

\begin{thebibliography}{65}%
\makeatletter
\providecommand \@ifxundefined [1]{%
 \@ifx{#1\undefined}
}%
\providecommand \@ifnum [1]{%
 \ifnum #1\expandafter \@firstoftwo
 \else \expandafter \@secondoftwo
 \fi
}%
\providecommand \@ifx [1]{%
 \ifx #1\expandafter \@firstoftwo
 \else \expandafter \@secondoftwo
 \fi
}%
\providecommand \natexlab [1]{#1}%
\providecommand \enquote  [1]{``#1''}%
\providecommand \bibnamefont  [1]{#1}%
\providecommand \bibfnamefont [1]{#1}%
\providecommand \citenamefont [1]{#1}%
\providecommand \href@noop [0]{\@secondoftwo}%
\providecommand \href [0]{\begingroup \@sanitize@url \@href}%
\providecommand \@href[1]{\@@startlink{#1}\@@href}%
\providecommand \@@href[1]{\endgroup#1\@@endlink}%
\providecommand \@sanitize@url [0]{\catcode `\\12\catcode `\$12\catcode `\&12\catcode `\#12\catcode `\^12\catcode `\_12\catcode `\%12\relax}%
\providecommand \@@startlink[1]{}%
\providecommand \@@endlink[0]{}%
\providecommand \url  [0]{\begingroup\@sanitize@url \@url }%
\providecommand \@url [1]{\endgroup\@href {#1}{\urlprefix }}%
\providecommand \urlprefix  [0]{URL }%
\providecommand \Eprint [0]{\href }%
\providecommand \doibase [0]{https://doi.org/}%
\providecommand \selectlanguage [0]{\@gobble}%
\providecommand \bibinfo  [0]{\@secondoftwo}%
\providecommand \bibfield  [0]{\@secondoftwo}%
\providecommand \translation [1]{[#1]}%
\providecommand \BibitemOpen [0]{}%
\providecommand \bibitemStop [0]{}%
\providecommand \bibitemNoStop [0]{.\EOS\space}%
\providecommand \EOS [0]{\spacefactor3000\relax}%
\providecommand \BibitemShut  [1]{\csname bibitem#1\endcsname}%
\let\auto@bib@innerbib\@empty
\bibitem [{\citenamefont {Bassett}\ and\ \citenamefont {Sporns}(2017)}]{bassett2017network}%
  \BibitemOpen
  \bibfield  {author} {\bibinfo {author} {\bibfnamefont {D.~S.}\ \bibnamefont {Bassett}}\ and\ \bibinfo {author} {\bibfnamefont {O.}~\bibnamefont {Sporns}},\ }\bibfield  {title} {\bibinfo {title} {Network neuroscience},\ }\href@noop {} {\bibfield  {journal} {\bibinfo  {journal} {Nature neuroscience}\ }\textbf {\bibinfo {volume} {20}},\ \bibinfo {pages} {353} (\bibinfo {year} {2017})}\BibitemShut {NoStop}%
\bibitem [{\citenamefont {Bashan}\ \emph {et~al.}(2012)\citenamefont {Bashan}, \citenamefont {Bartsch}, \citenamefont {Kantelhardt}, \citenamefont {Havlin},\ and\ \citenamefont {Ivanov}}]{bashan2012network}%
  \BibitemOpen
  \bibfield  {author} {\bibinfo {author} {\bibfnamefont {A.}~\bibnamefont {Bashan}}, \bibinfo {author} {\bibfnamefont {R.~P.}\ \bibnamefont {Bartsch}}, \bibinfo {author} {\bibfnamefont {J.~W.}\ \bibnamefont {Kantelhardt}}, \bibinfo {author} {\bibfnamefont {S.}~\bibnamefont {Havlin}},\ and\ \bibinfo {author} {\bibfnamefont {P.~C.}\ \bibnamefont {Ivanov}},\ }\bibfield  {title} {\bibinfo {title} {Network physiology reveals relations between network topology and physiological function},\ }\href@noop {} {\bibfield  {journal} {\bibinfo  {journal} {Nature communications}\ }\textbf {\bibinfo {volume} {3}},\ \bibinfo {pages} {702} (\bibinfo {year} {2012})}\BibitemShut {NoStop}%
\bibitem [{\citenamefont {Barab{\'a}si}(2013)}]{barabasi2013network}%
  \BibitemOpen
  \bibfield  {author} {\bibinfo {author} {\bibfnamefont {A.-L.}\ \bibnamefont {Barab{\'a}si}},\ }\bibfield  {title} {\bibinfo {title} {Network science},\ }\href@noop {} {\bibfield  {journal} {\bibinfo  {journal} {Philosophical Transactions of the Royal Society A: Mathematical, Physical and Engineering Sciences}\ }\textbf {\bibinfo {volume} {371}},\ \bibinfo {pages} {20120375} (\bibinfo {year} {2013})}\BibitemShut {NoStop}%
\bibitem [{\citenamefont {Rubinov}\ and\ \citenamefont {Sporns}(2010)}]{rubinov2010complex}%
  \BibitemOpen
  \bibfield  {author} {\bibinfo {author} {\bibfnamefont {M.}~\bibnamefont {Rubinov}}\ and\ \bibinfo {author} {\bibfnamefont {O.}~\bibnamefont {Sporns}},\ }\bibfield  {title} {\bibinfo {title} {Complex network measures of brain connectivity: uses and interpretations},\ }\href@noop {} {\bibfield  {journal} {\bibinfo  {journal} {Neuroimage}\ }\textbf {\bibinfo {volume} {52}},\ \bibinfo {pages} {1059} (\bibinfo {year} {2010})}\BibitemShut {NoStop}%
\bibitem [{\citenamefont {Lehnertz}\ \emph {et~al.}(2020)\citenamefont {Lehnertz}, \citenamefont {Br{\"o}hl},\ and\ \citenamefont {Rings}}]{lehnertz2020human}%
  \BibitemOpen
  \bibfield  {author} {\bibinfo {author} {\bibfnamefont {K.}~\bibnamefont {Lehnertz}}, \bibinfo {author} {\bibfnamefont {T.}~\bibnamefont {Br{\"o}hl}},\ and\ \bibinfo {author} {\bibfnamefont {T.}~\bibnamefont {Rings}},\ }\bibfield  {title} {\bibinfo {title} {The human organism as an integrated interaction network: Recent conceptual and methodological challenges},\ }\href@noop {} {\bibfield  {journal} {\bibinfo  {journal} {Frontiers in Physiology}\ }\textbf {\bibinfo {volume} {11}},\ \bibinfo {pages} {598694} (\bibinfo {year} {2020})}\BibitemShut {NoStop}%
\bibitem [{\citenamefont {Butts}(2009)}]{butts2009revisiting}%
  \BibitemOpen
  \bibfield  {author} {\bibinfo {author} {\bibfnamefont {C.~T.}\ \bibnamefont {Butts}},\ }\bibfield  {title} {\bibinfo {title} {Revisiting the foundations of network analysis},\ }\href@noop {} {\bibfield  {journal} {\bibinfo  {journal} {science}\ }\textbf {\bibinfo {volume} {325}},\ \bibinfo {pages} {414} (\bibinfo {year} {2009})}\BibitemShut {NoStop}%
\bibitem [{\citenamefont {Pincus}\ and\ \citenamefont {Goldberger}(1994)}]{pincus1994physiological}%
  \BibitemOpen
  \bibfield  {author} {\bibinfo {author} {\bibfnamefont {S.~M.}\ \bibnamefont {Pincus}}\ and\ \bibinfo {author} {\bibfnamefont {A.~L.}\ \bibnamefont {Goldberger}},\ }\bibfield  {title} {\bibinfo {title} {Physiological time-series analysis: what does regularity quantify?},\ }\href@noop {} {\bibfield  {journal} {\bibinfo  {journal} {American Journal of Physiology-Heart and Circulatory Physiology}\ }\textbf {\bibinfo {volume} {266}},\ \bibinfo {pages} {H1643} (\bibinfo {year} {1994})}\BibitemShut {NoStop}%
\bibitem [{\citenamefont {Pereda}\ \emph {et~al.}(2005)\citenamefont {Pereda}, \citenamefont {Quiroga},\ and\ \citenamefont {Bhattacharya}}]{pereda2005nonlinear}%
  \BibitemOpen
  \bibfield  {author} {\bibinfo {author} {\bibfnamefont {E.}~\bibnamefont {Pereda}}, \bibinfo {author} {\bibfnamefont {R.~Q.}\ \bibnamefont {Quiroga}},\ and\ \bibinfo {author} {\bibfnamefont {J.}~\bibnamefont {Bhattacharya}},\ }\bibfield  {title} {\bibinfo {title} {Nonlinear multivariate analysis of neurophysiological signals},\ }\href@noop {} {\bibfield  {journal} {\bibinfo  {journal} {Progress in neurobiology}\ }\textbf {\bibinfo {volume} {77}},\ \bibinfo {pages} {1} (\bibinfo {year} {2005})}\BibitemShut {NoStop}%
\bibitem [{\citenamefont {Battiston}\ \emph {et~al.}(2020)\citenamefont {Battiston}, \citenamefont {Cencetti}, \citenamefont {Iacopini}, \citenamefont {Latora}, \citenamefont {Lucas}, \citenamefont {Patania}, \citenamefont {Young},\ and\ \citenamefont {Petri}}]{battiston2020networks}%
  \BibitemOpen
  \bibfield  {author} {\bibinfo {author} {\bibfnamefont {F.}~\bibnamefont {Battiston}}, \bibinfo {author} {\bibfnamefont {G.}~\bibnamefont {Cencetti}}, \bibinfo {author} {\bibfnamefont {I.}~\bibnamefont {Iacopini}}, \bibinfo {author} {\bibfnamefont {V.}~\bibnamefont {Latora}}, \bibinfo {author} {\bibfnamefont {M.}~\bibnamefont {Lucas}}, \bibinfo {author} {\bibfnamefont {A.}~\bibnamefont {Patania}}, \bibinfo {author} {\bibfnamefont {J.-G.}\ \bibnamefont {Young}},\ and\ \bibinfo {author} {\bibfnamefont {G.}~\bibnamefont {Petri}},\ }\bibfield  {title} {\bibinfo {title} {Networks beyond pairwise interactions: Structure and dynamics},\ }\href@noop {} {\bibfield  {journal} {\bibinfo  {journal} {Physics Reports}\ }\textbf {\bibinfo {volume} {874}},\ \bibinfo {pages} {1} (\bibinfo {year} {2020})}\BibitemShut {NoStop}%
\bibitem [{\citenamefont {Stramaglia}\ \emph {et~al.}(2014)\citenamefont {Stramaglia}, \citenamefont {Cortes},\ and\ \citenamefont {Marinazzo}}]{stramaglia2014synergy}%
  \BibitemOpen
  \bibfield  {author} {\bibinfo {author} {\bibfnamefont {S.}~\bibnamefont {Stramaglia}}, \bibinfo {author} {\bibfnamefont {J.~M.}\ \bibnamefont {Cortes}},\ and\ \bibinfo {author} {\bibfnamefont {D.}~\bibnamefont {Marinazzo}},\ }\bibfield  {title} {\bibinfo {title} {Synergy and redundancy in the granger causal analysis of dynamical networks},\ }\href@noop {} {\bibfield  {journal} {\bibinfo  {journal} {New Journal of Physics}\ }\textbf {\bibinfo {volume} {16}},\ \bibinfo {pages} {105003} (\bibinfo {year} {2014})}\BibitemShut {NoStop}%
\bibitem [{\citenamefont {Faes}\ \emph {et~al.}(2016)\citenamefont {Faes}, \citenamefont {Porta}, \citenamefont {Nollo},\ and\ \citenamefont {Javorka}}]{faes2016information}%
  \BibitemOpen
  \bibfield  {author} {\bibinfo {author} {\bibfnamefont {L.}~\bibnamefont {Faes}}, \bibinfo {author} {\bibfnamefont {A.}~\bibnamefont {Porta}}, \bibinfo {author} {\bibfnamefont {G.}~\bibnamefont {Nollo}},\ and\ \bibinfo {author} {\bibfnamefont {M.}~\bibnamefont {Javorka}},\ }\bibfield  {title} {\bibinfo {title} {Information decomposition in multivariate systems: definitions, implementation and application to cardiovascular networks},\ }\href@noop {} {\bibfield  {journal} {\bibinfo  {journal} {Entropy}\ }\textbf {\bibinfo {volume} {19}},\ \bibinfo {pages} {5} (\bibinfo {year} {2016})}\BibitemShut {NoStop}%
\bibitem [{\citenamefont {Courtney}\ and\ \citenamefont {Bianconi}(2016)}]{courtney2016generalized}%
  \BibitemOpen
  \bibfield  {author} {\bibinfo {author} {\bibfnamefont {O.~T.}\ \bibnamefont {Courtney}}\ and\ \bibinfo {author} {\bibfnamefont {G.}~\bibnamefont {Bianconi}},\ }\bibfield  {title} {\bibinfo {title} {Generalized network structures: The configuration model and the canonical ensemble of simplicial complexes},\ }\href@noop {} {\bibfield  {journal} {\bibinfo  {journal} {Physical Review E}\ }\textbf {\bibinfo {volume} {93}},\ \bibinfo {pages} {062311} (\bibinfo {year} {2016})}\BibitemShut {NoStop}%
\bibitem [{\citenamefont {Wibral}\ \emph {et~al.}(2014)\citenamefont {Wibral}, \citenamefont {Vicente},\ and\ \citenamefont {Lizier}}]{wibral2014directed}%
  \BibitemOpen
  \bibfield  {author} {\bibinfo {author} {\bibfnamefont {M.}~\bibnamefont {Wibral}}, \bibinfo {author} {\bibfnamefont {R.}~\bibnamefont {Vicente}},\ and\ \bibinfo {author} {\bibfnamefont {J.~T.}\ \bibnamefont {Lizier}},\ }\href@noop {} {\emph {\bibinfo {title} {Directed information measures in neuroscience}}},\ Vol.\ \bibinfo {volume} {724}\ (\bibinfo  {publisher} {Springer},\ \bibinfo {year} {2014})\BibitemShut {NoStop}%
\bibitem [{\citenamefont {James}\ \emph {et~al.}(2016)\citenamefont {James}, \citenamefont {Barnett},\ and\ \citenamefont {Crutchfield}}]{james2016information}%
  \BibitemOpen
  \bibfield  {author} {\bibinfo {author} {\bibfnamefont {R.~G.}\ \bibnamefont {James}}, \bibinfo {author} {\bibfnamefont {N.}~\bibnamefont {Barnett}},\ and\ \bibinfo {author} {\bibfnamefont {J.~P.}\ \bibnamefont {Crutchfield}},\ }\bibfield  {title} {\bibinfo {title} {Information flows? a critique of transfer entropies},\ }\href@noop {} {\bibfield  {journal} {\bibinfo  {journal} {Physical review letters}\ }\textbf {\bibinfo {volume} {116}},\ \bibinfo {pages} {238701} (\bibinfo {year} {2016})}\BibitemShut {NoStop}%
\bibitem [{\citenamefont {Lizier}\ \emph {et~al.}(2018)\citenamefont {Lizier}, \citenamefont {Bertschinger}, \citenamefont {Jost},\ and\ \citenamefont {Wibral}}]{lizier2018information}%
  \BibitemOpen
  \bibfield  {author} {\bibinfo {author} {\bibfnamefont {J.~T.}\ \bibnamefont {Lizier}}, \bibinfo {author} {\bibfnamefont {N.}~\bibnamefont {Bertschinger}}, \bibinfo {author} {\bibfnamefont {J.}~\bibnamefont {Jost}},\ and\ \bibinfo {author} {\bibfnamefont {M.}~\bibnamefont {Wibral}},\ }\href@noop {} {\bibinfo {title} {Information decomposition of target effects from multi-source interactions: Perspectives on previous, current and future work}} (\bibinfo {year} {2018})\BibitemShut {NoStop}%
\bibitem [{\citenamefont {Faes}\ \emph {et~al.}(2021)\citenamefont {Faes}, \citenamefont {Pernice}, \citenamefont {Mijatovic}, \citenamefont {Antonacci}, \citenamefont {Krohova}, \citenamefont {Javorka},\ and\ \citenamefont {Porta}}]{faes2021information}%
  \BibitemOpen
  \bibfield  {author} {\bibinfo {author} {\bibfnamefont {L.}~\bibnamefont {Faes}}, \bibinfo {author} {\bibfnamefont {R.}~\bibnamefont {Pernice}}, \bibinfo {author} {\bibfnamefont {G.}~\bibnamefont {Mijatovic}}, \bibinfo {author} {\bibfnamefont {Y.}~\bibnamefont {Antonacci}}, \bibinfo {author} {\bibfnamefont {J.~C.}\ \bibnamefont {Krohova}}, \bibinfo {author} {\bibfnamefont {M.}~\bibnamefont {Javorka}},\ and\ \bibinfo {author} {\bibfnamefont {A.}~\bibnamefont {Porta}},\ }\bibfield  {title} {\bibinfo {title} {Information decomposition in the frequency domain: A new framework to study cardiovascular and cardiorespiratory oscillations},\ }\href@noop {} {\bibfield  {journal} {\bibinfo  {journal} {Philosophical Transactions of the Royal Society A}\ }\textbf {\bibinfo {volume} {379}},\ \bibinfo {pages} {20200250} (\bibinfo {year} {2021})}\BibitemShut {NoStop}%
\bibitem [{\citenamefont {McGill}(1954)}]{mcgill1954multivariate}%
  \BibitemOpen
  \bibfield  {author} {\bibinfo {author} {\bibfnamefont {W.}~\bibnamefont {McGill}},\ }\bibfield  {title} {\bibinfo {title} {Multivariate information transmission},\ }\href@noop {} {\bibfield  {journal} {\bibinfo  {journal} {Transactions of the IRE Professional Group on Information Theory}\ }\textbf {\bibinfo {volume} {4}},\ \bibinfo {pages} {93} (\bibinfo {year} {1954})}\BibitemShut {NoStop}%
\bibitem [{\citenamefont {Gelfand}\ and\ \citenamefont {Iaglom}(1959)}]{gelfand1959calculation}%
  \BibitemOpen
  \bibfield  {author} {\bibinfo {author} {\bibfnamefont {I.~M.}\ \bibnamefont {Gelfand}}\ and\ \bibinfo {author} {\bibfnamefont {A.}~\bibnamefont {Iaglom}},\ }\href@noop {} {\emph {\bibinfo {title} {Calculation of the amount of information about a random function contained in another such function}}}\ (\bibinfo  {publisher} {American Mathematical Society Providence},\ \bibinfo {year} {1959})\BibitemShut {NoStop}%
\bibitem [{\citenamefont {Kolmogorov}(1959)}]{kolmogorov1959entropy}%
  \BibitemOpen
  \bibfield  {author} {\bibinfo {author} {\bibfnamefont {A.~N.}\ \bibnamefont {Kolmogorov}},\ }\bibfield  {title} {\bibinfo {title} {Entropy per unit time as a metric invariant of automorphisms},\ }in\ \href@noop {} {\emph {\bibinfo {booktitle} {Dokl. Akad. Nauk SSSR}}},\ Vol.\ \bibinfo {volume} {124}\ (\bibinfo {year} {1959})\ pp.\ \bibinfo {pages} {754--755}\BibitemShut {NoStop}%
\bibitem [{\citenamefont {Duncan}(1970)}]{duncan1970calculation}%
  \BibitemOpen
  \bibfield  {author} {\bibinfo {author} {\bibfnamefont {T.~E.}\ \bibnamefont {Duncan}},\ }\bibfield  {title} {\bibinfo {title} {On the calculation of mutual information},\ }\href@noop {} {\bibfield  {journal} {\bibinfo  {journal} {SIAM Journal on Applied Mathematics}\ }\textbf {\bibinfo {volume} {19}},\ \bibinfo {pages} {215} (\bibinfo {year} {1970})}\BibitemShut {NoStop}%
\bibitem [{\citenamefont {Cover}(1999)}]{cover1999elements}%
  \BibitemOpen
  \bibfield  {author} {\bibinfo {author} {\bibfnamefont {T.~M.}\ \bibnamefont {Cover}},\ }\href@noop {} {\emph {\bibinfo {title} {Elements of information theory}}}\ (\bibinfo  {publisher} {John Wiley \& Sons},\ \bibinfo {year} {1999})\BibitemShut {NoStop}%
\bibitem [{\citenamefont {Chicharro}(2011)}]{chicharro2011spectral}%
  \BibitemOpen
  \bibfield  {author} {\bibinfo {author} {\bibfnamefont {D.}~\bibnamefont {Chicharro}},\ }\bibfield  {title} {\bibinfo {title} {On the spectral formulation of granger causality},\ }\href@noop {} {\bibfield  {journal} {\bibinfo  {journal} {Biological cybernetics}\ }\textbf {\bibinfo {volume} {105}},\ \bibinfo {pages} {331} (\bibinfo {year} {2011})}\BibitemShut {NoStop}%
\bibitem [{\citenamefont {Rosas}\ \emph {et~al.}(2019)\citenamefont {Rosas}, \citenamefont {Mediano}, \citenamefont {Gastpar},\ and\ \citenamefont {Jensen}}]{rosas2019quantifying}%
  \BibitemOpen
  \bibfield  {author} {\bibinfo {author} {\bibfnamefont {F.~E.}\ \bibnamefont {Rosas}}, \bibinfo {author} {\bibfnamefont {P.~A.}\ \bibnamefont {Mediano}}, \bibinfo {author} {\bibfnamefont {M.}~\bibnamefont {Gastpar}},\ and\ \bibinfo {author} {\bibfnamefont {H.~J.}\ \bibnamefont {Jensen}},\ }\bibfield  {title} {\bibinfo {title} {Quantifying high-order interdependencies via multivariate extensions of the mutual information},\ }\href@noop {} {\bibfield  {journal} {\bibinfo  {journal} {Physical Review E}\ }\textbf {\bibinfo {volume} {100}},\ \bibinfo {pages} {032305} (\bibinfo {year} {2019})}\BibitemShut {NoStop}%
\bibitem [{\citenamefont {Antonacci}\ \emph {et~al.}(2021)\citenamefont {Antonacci}, \citenamefont {Minati}, \citenamefont {Nuzzi}, \citenamefont {Mijatovic}, \citenamefont {Pernice}, \citenamefont {Marinazzo}, \citenamefont {Stramaglia},\ and\ \citenamefont {Faes}}]{antonacci2021measuring}%
  \BibitemOpen
  \bibfield  {author} {\bibinfo {author} {\bibfnamefont {Y.}~\bibnamefont {Antonacci}}, \bibinfo {author} {\bibfnamefont {L.}~\bibnamefont {Minati}}, \bibinfo {author} {\bibfnamefont {D.}~\bibnamefont {Nuzzi}}, \bibinfo {author} {\bibfnamefont {G.}~\bibnamefont {Mijatovic}}, \bibinfo {author} {\bibfnamefont {R.}~\bibnamefont {Pernice}}, \bibinfo {author} {\bibfnamefont {D.}~\bibnamefont {Marinazzo}}, \bibinfo {author} {\bibfnamefont {S.}~\bibnamefont {Stramaglia}},\ and\ \bibinfo {author} {\bibfnamefont {L.}~\bibnamefont {Faes}},\ }\bibfield  {title} {\bibinfo {title} {Measuring high-order interactions in rhythmic processes through multivariate spectral information decomposition},\ }\href@noop {} {\bibfield  {journal} {\bibinfo  {journal} {IEEE Access}\ }\textbf {\bibinfo {volume} {9}},\ \bibinfo {pages} {149486} (\bibinfo {year} {2021})}\BibitemShut {NoStop}%
\bibitem [{\citenamefont {Faes}\ \emph {et~al.}(2022{\natexlab{a}})\citenamefont {Faes}, \citenamefont {Mijatovic}, \citenamefont {Antonacci}, \citenamefont {Pernice}, \citenamefont {Bara}, \citenamefont {Sparacino}, \citenamefont {Sammartino}, \citenamefont {Porta}, \citenamefont {Marinazzo},\ and\ \citenamefont {Stramaglia}}]{faes2022new}%
  \BibitemOpen
  \bibfield  {author} {\bibinfo {author} {\bibfnamefont {L.}~\bibnamefont {Faes}}, \bibinfo {author} {\bibfnamefont {G.}~\bibnamefont {Mijatovic}}, \bibinfo {author} {\bibfnamefont {Y.}~\bibnamefont {Antonacci}}, \bibinfo {author} {\bibfnamefont {R.}~\bibnamefont {Pernice}}, \bibinfo {author} {\bibfnamefont {C.}~\bibnamefont {Bara}}, \bibinfo {author} {\bibfnamefont {L.}~\bibnamefont {Sparacino}}, \bibinfo {author} {\bibfnamefont {M.}~\bibnamefont {Sammartino}}, \bibinfo {author} {\bibfnamefont {A.}~\bibnamefont {Porta}}, \bibinfo {author} {\bibfnamefont {D.}~\bibnamefont {Marinazzo}},\ and\ \bibinfo {author} {\bibfnamefont {S.}~\bibnamefont {Stramaglia}},\ }\bibfield  {title} {\bibinfo {title} {A new framework for the time-and frequency-domain assessment of high-order interactions in networks of random processes},\ }\href@noop {} {\bibfield  {journal} {\bibinfo  {journal} {IEEE Transactions on Signal Processing}\ }\textbf {\bibinfo {volume} {70}},\ \bibinfo {pages} {5766} (\bibinfo {year}
  {2022}{\natexlab{a}})}\BibitemShut {NoStop}%
\bibitem [{\citenamefont {Bara}\ \emph {et~al.}(2023)\citenamefont {Bara}, \citenamefont {Sparacino}, \citenamefont {Pernice}, \citenamefont {Antonacci}, \citenamefont {Porta}, \citenamefont {Kugiumtzis},\ and\ \citenamefont {Faes}}]{bara2023comparison}%
  \BibitemOpen
  \bibfield  {author} {\bibinfo {author} {\bibfnamefont {C.}~\bibnamefont {Bara}}, \bibinfo {author} {\bibfnamefont {L.}~\bibnamefont {Sparacino}}, \bibinfo {author} {\bibfnamefont {R.}~\bibnamefont {Pernice}}, \bibinfo {author} {\bibfnamefont {Y.}~\bibnamefont {Antonacci}}, \bibinfo {author} {\bibfnamefont {A.}~\bibnamefont {Porta}}, \bibinfo {author} {\bibfnamefont {D.}~\bibnamefont {Kugiumtzis}},\ and\ \bibinfo {author} {\bibfnamefont {L.}~\bibnamefont {Faes}},\ }\bibfield  {title} {\bibinfo {title} {Comparison of discretization strategies for the model-free information-theoretic assessment of short-term physiological interactions},\ }\href@noop {} {\bibfield  {journal} {\bibinfo  {journal} {Chaos: An Interdisciplinary Journal of Nonlinear Science}\ }\textbf {\bibinfo {volume} {33}} (\bibinfo {year} {2023})}\BibitemShut {NoStop}%
\bibitem [{\citenamefont {Shannon}(1948)}]{shannon1948mathematical}%
  \BibitemOpen
  \bibfield  {author} {\bibinfo {author} {\bibfnamefont {C.~E.}\ \bibnamefont {Shannon}},\ }\bibfield  {title} {\bibinfo {title} {A mathematical theory of communication},\ }\href@noop {} {\bibfield  {journal} {\bibinfo  {journal} {The Bell system technical journal}\ }\textbf {\bibinfo {volume} {27}},\ \bibinfo {pages} {379} (\bibinfo {year} {1948})}\BibitemShut {NoStop}%
\bibitem [{\citenamefont {Geweke}(1982)}]{geweke1982measurement}%
  \BibitemOpen
  \bibfield  {author} {\bibinfo {author} {\bibfnamefont {J.}~\bibnamefont {Geweke}},\ }\bibfield  {title} {\bibinfo {title} {Measurement of linear dependence and feedback between multiple time series},\ }\href@noop {} {\bibfield  {journal} {\bibinfo  {journal} {Journal of the American statistical association}\ }\textbf {\bibinfo {volume} {77}},\ \bibinfo {pages} {304} (\bibinfo {year} {1982})}\BibitemShut {NoStop}%
\bibitem [{\citenamefont {Kay}(1988)}]{kay1988modern}%
  \BibitemOpen
  \bibfield  {author} {\bibinfo {author} {\bibfnamefont {S.~M.}\ \bibnamefont {Kay}},\ }\href@noop {} {\emph {\bibinfo {title} {Modern spectral estimation: theory and application}}}\ (\bibinfo  {publisher} {Pearson Education India},\ \bibinfo {year} {1988})\BibitemShut {NoStop}%
\bibitem [{\citenamefont {Pinna}\ \emph {et~al.}(1996)\citenamefont {Pinna}, \citenamefont {Maestri},\ and\ \citenamefont {Di~Cesare}}]{pinna1996application}%
  \BibitemOpen
  \bibfield  {author} {\bibinfo {author} {\bibfnamefont {G.}~\bibnamefont {Pinna}}, \bibinfo {author} {\bibfnamefont {R.}~\bibnamefont {Maestri}},\ and\ \bibinfo {author} {\bibfnamefont {A.}~\bibnamefont {Di~Cesare}},\ }\bibfield  {title} {\bibinfo {title} {Application of time series spectral analysis theory: analysis of cardiovascular variability signals},\ }\href@noop {} {\bibfield  {journal} {\bibinfo  {journal} {Medical and Biological Engineering and Computing}\ }\textbf {\bibinfo {volume} {34}},\ \bibinfo {pages} {142} (\bibinfo {year} {1996})}\BibitemShut {NoStop}%
\bibitem [{\citenamefont {Zhao}\ and\ \citenamefont {Gui}(2019)}]{zhao2019nonparametric}%
  \BibitemOpen
  \bibfield  {author} {\bibinfo {author} {\bibfnamefont {H.}~\bibnamefont {Zhao}}\ and\ \bibinfo {author} {\bibfnamefont {L.}~\bibnamefont {Gui}},\ }\bibfield  {title} {\bibinfo {title} {Nonparametric and parametric methods of spectral analysis},\ }in\ \href@noop {} {\emph {\bibinfo {booktitle} {Matec web of conferences}}},\ Vol.\ \bibinfo {volume} {283}\ (\bibinfo {organization} {EDP Sciences},\ \bibinfo {year} {2019})\ p.\ \bibinfo {pages} {07002}\BibitemShut {NoStop}%
\bibitem [{\citenamefont {Blackman}\ and\ \citenamefont {Tukey}(1958)}]{blackman1958measurement}%
  \BibitemOpen
  \bibfield  {author} {\bibinfo {author} {\bibfnamefont {R.~B.}\ \bibnamefont {Blackman}}\ and\ \bibinfo {author} {\bibfnamefont {J.~W.}\ \bibnamefont {Tukey}},\ }\bibfield  {title} {\bibinfo {title} {The measurement of power spectra from the point of view of communications engineering—part i},\ }\href@noop {} {\bibfield  {journal} {\bibinfo  {journal} {Bell System Technical Journal}\ }\textbf {\bibinfo {volume} {37}},\ \bibinfo {pages} {185} (\bibinfo {year} {1958})}\BibitemShut {NoStop}%
\bibitem [{\citenamefont {Priestley}(1981)}]{priestley1981spectral}%
  \BibitemOpen
  \bibfield  {author} {\bibinfo {author} {\bibfnamefont {M.~B.}\ \bibnamefont {Priestley}},\ }\href@noop {} {\emph {\bibinfo {title} {Spectral analysis and time series: probability and mathematical statistics}}},\ \bibinfo {number} {04; QA280, P7.}\ (\bibinfo {year} {1981})\BibitemShut {NoStop}%
\bibitem [{\citenamefont {L{\"u}tkepohl}(2005)}]{lutkepohl2005new}%
  \BibitemOpen
  \bibfield  {author} {\bibinfo {author} {\bibfnamefont {H.}~\bibnamefont {L{\"u}tkepohl}},\ }\href@noop {} {\emph {\bibinfo {title} {New introduction to multiple time series analysis}}}\ (\bibinfo  {publisher} {Springer Science \& Business Media},\ \bibinfo {year} {2005})\BibitemShut {NoStop}%
\bibitem [{\citenamefont {Schl{\"o}gl}(2006)}]{schlogl2006comparison}%
  \BibitemOpen
  \bibfield  {author} {\bibinfo {author} {\bibfnamefont {A.}~\bibnamefont {Schl{\"o}gl}},\ }\bibfield  {title} {\bibinfo {title} {A comparison of multivariate autoregressive estimators},\ }\href@noop {} {\bibfield  {journal} {\bibinfo  {journal} {Signal processing}\ }\textbf {\bibinfo {volume} {86}},\ \bibinfo {pages} {2426} (\bibinfo {year} {2006})}\BibitemShut {NoStop}%
\bibitem [{\citenamefont {Antonacci}\ \emph {et~al.}(2020)\citenamefont {Antonacci}, \citenamefont {Astolfi}, \citenamefont {Nollo},\ and\ \citenamefont {Faes}}]{antonacci2020information}%
  \BibitemOpen
  \bibfield  {author} {\bibinfo {author} {\bibfnamefont {Y.}~\bibnamefont {Antonacci}}, \bibinfo {author} {\bibfnamefont {L.}~\bibnamefont {Astolfi}}, \bibinfo {author} {\bibfnamefont {G.}~\bibnamefont {Nollo}},\ and\ \bibinfo {author} {\bibfnamefont {L.}~\bibnamefont {Faes}},\ }\bibfield  {title} {\bibinfo {title} {Information transfer in linear multivariate processes assessed through penalized regression techniques: validation and application to physiological networks},\ }\href@noop {} {\bibfield  {journal} {\bibinfo  {journal} {Entropy}\ }\textbf {\bibinfo {volume} {22}},\ \bibinfo {pages} {732} (\bibinfo {year} {2020})}\BibitemShut {NoStop}%
\bibitem [{\citenamefont {Karimi}(2011)}]{karimi2011order}%
  \BibitemOpen
  \bibfield  {author} {\bibinfo {author} {\bibfnamefont {M.}~\bibnamefont {Karimi}},\ }\bibfield  {title} {\bibinfo {title} {Order selection criteria for vector autoregressive models},\ }\href@noop {} {\bibfield  {journal} {\bibinfo  {journal} {Signal Processing}\ }\textbf {\bibinfo {volume} {91}},\ \bibinfo {pages} {955} (\bibinfo {year} {2011})}\BibitemShut {NoStop}%
\bibitem [{\citenamefont {Akaike}(1974)}]{akaike1974new}%
  \BibitemOpen
  \bibfield  {author} {\bibinfo {author} {\bibfnamefont {H.}~\bibnamefont {Akaike}},\ }\bibfield  {title} {\bibinfo {title} {A new look at the statistical model identification},\ }\href@noop {} {\bibfield  {journal} {\bibinfo  {journal} {IEEE transactions on automatic control}\ }\textbf {\bibinfo {volume} {19}},\ \bibinfo {pages} {716} (\bibinfo {year} {1974})}\BibitemShut {NoStop}%
\bibitem [{\citenamefont {Schwarz}(1978)}]{schwarz1978estimating}%
  \BibitemOpen
  \bibfield  {author} {\bibinfo {author} {\bibfnamefont {G.}~\bibnamefont {Schwarz}},\ }\bibfield  {title} {\bibinfo {title} {Estimating the dimension of a model},\ }\href@noop {} {\bibfield  {journal} {\bibinfo  {journal} {The annals of statistics}\ ,\ \bibinfo {pages} {461}} (\bibinfo {year} {1978})}\BibitemShut {NoStop}%
\bibitem [{\citenamefont {Theiler}\ \emph {et~al.}(1992)\citenamefont {Theiler}, \citenamefont {Eubank}, \citenamefont {Longtin}, \citenamefont {Galdrikian},\ and\ \citenamefont {Farmer}}]{theiler1992}%
  \BibitemOpen
  \bibfield  {author} {\bibinfo {author} {\bibfnamefont {J.}~\bibnamefont {Theiler}}, \bibinfo {author} {\bibfnamefont {S.}~\bibnamefont {Eubank}}, \bibinfo {author} {\bibfnamefont {A.}~\bibnamefont {Longtin}}, \bibinfo {author} {\bibfnamefont {B.}~\bibnamefont {Galdrikian}},\ and\ \bibinfo {author} {\bibfnamefont {J.~D.}\ \bibnamefont {Farmer}},\ }\bibfield  {title} {\bibinfo {title} {Testing for nonlinearity in time series: the method of surrogate data},\ }\href@noop {} {\bibfield  {journal} {\bibinfo  {journal} {Physica D: Nonlinear Phenomena}\ }\textbf {\bibinfo {volume} {58}},\ \bibinfo {pages} {77} (\bibinfo {year} {1992})}\BibitemShut {NoStop}%
\bibitem [{\citenamefont {Palus}(1997)}]{palus1997}%
  \BibitemOpen
  \bibfield  {author} {\bibinfo {author} {\bibfnamefont {M.}~\bibnamefont {Palus}},\ }\bibfield  {title} {\bibinfo {title} {Detecting phase synchronization in noisy systems},\ }\href@noop {} {\bibfield  {journal} {\bibinfo  {journal} {Physics Letters A}\ }\textbf {\bibinfo {volume} {235}},\ \bibinfo {pages} {341} (\bibinfo {year} {1997})}\BibitemShut {NoStop}%
\bibitem [{\citenamefont {Schreiber}\ and\ \citenamefont {Schmitz}(1996)}]{schreiber1996surrogate}%
  \BibitemOpen
  \bibfield  {author} {\bibinfo {author} {\bibfnamefont {T.}~\bibnamefont {Schreiber}}\ and\ \bibinfo {author} {\bibfnamefont {A.}~\bibnamefont {Schmitz}},\ }\bibfield  {title} {\bibinfo {title} {Improved surrogate data for nonlinearity tests},\ }\href@noop {} {\bibfield  {journal} {\bibinfo  {journal} {Physical review letters}\ }\textbf {\bibinfo {volume} {77}},\ \bibinfo {pages} {635} (\bibinfo {year} {1996})}\BibitemShut {NoStop}%
\bibitem [{\citenamefont {Efron}(1979)}]{efron1979bootstrap_first}%
  \BibitemOpen
  \bibfield  {author} {\bibinfo {author} {\bibfnamefont {B.}~\bibnamefont {Efron}},\ }\bibfield  {title} {\bibinfo {title} {Bootstrap methods: another look at the jackknife},\ }\href@noop {} {\bibfield  {journal} {\bibinfo  {journal} {The Annals of Statistics}\ }\textbf {\bibinfo {volume} {7}},\ \bibinfo {pages} {1} (\bibinfo {year} {1979})}\BibitemShut {NoStop}%
\bibitem [{\citenamefont {Politis}(2003)}]{politis2003bootstrap}%
  \BibitemOpen
  \bibfield  {author} {\bibinfo {author} {\bibfnamefont {D.~N.}\ \bibnamefont {Politis}},\ }\bibfield  {title} {\bibinfo {title} {The impact of bootstrap methods on time series analysis},\ }\href@noop {} {\bibfield  {journal} {\bibinfo  {journal} {Statistical Science}\ ,\ \bibinfo {pages} {219}} (\bibinfo {year} {2003})}\BibitemShut {NoStop}%
\bibitem [{\citenamefont {Sparacino}\ \emph {et~al.}(2023)\citenamefont {Sparacino}, \citenamefont {Faes}, \citenamefont {Mijatovi{\'c}}, \citenamefont {Parla}, \citenamefont {Lo~Re}, \citenamefont {Miraglia}, \citenamefont {de~Ville~de Goyet},\ and\ \citenamefont {Sparacia}}]{sparacino2023statistical}%
  \BibitemOpen
  \bibfield  {author} {\bibinfo {author} {\bibfnamefont {L.}~\bibnamefont {Sparacino}}, \bibinfo {author} {\bibfnamefont {L.}~\bibnamefont {Faes}}, \bibinfo {author} {\bibfnamefont {G.}~\bibnamefont {Mijatovi{\'c}}}, \bibinfo {author} {\bibfnamefont {G.}~\bibnamefont {Parla}}, \bibinfo {author} {\bibfnamefont {V.}~\bibnamefont {Lo~Re}}, \bibinfo {author} {\bibfnamefont {R.}~\bibnamefont {Miraglia}}, \bibinfo {author} {\bibfnamefont {J.}~\bibnamefont {de~Ville~de Goyet}},\ and\ \bibinfo {author} {\bibfnamefont {G.}~\bibnamefont {Sparacia}},\ }\bibfield  {title} {\bibinfo {title} {Statistical approaches to identify pairwise and high-order brain functional connectivity signatures on a single-subject basis},\ }\href@noop {} {\bibfield  {journal} {\bibinfo  {journal} {Life}\ }\textbf {\bibinfo {volume} {13}},\ \bibinfo {pages} {2075} (\bibinfo {year} {2023})}\BibitemShut {NoStop}%
\bibitem [{\citenamefont {Faes}\ \emph {et~al.}(2015)\citenamefont {Faes}, \citenamefont {Porta},\ and\ \citenamefont {Nollo}}]{faes2015information}%
  \BibitemOpen
  \bibfield  {author} {\bibinfo {author} {\bibfnamefont {L.}~\bibnamefont {Faes}}, \bibinfo {author} {\bibfnamefont {A.}~\bibnamefont {Porta}},\ and\ \bibinfo {author} {\bibfnamefont {G.}~\bibnamefont {Nollo}},\ }\bibfield  {title} {\bibinfo {title} {Information decomposition in bivariate systems: theory and application to cardiorespiratory dynamics},\ }\href@noop {} {\bibfield  {journal} {\bibinfo  {journal} {Entropy}\ }\textbf {\bibinfo {volume} {17}},\ \bibinfo {pages} {277} (\bibinfo {year} {2015})}\BibitemShut {NoStop}%
\bibitem [{\citenamefont {Faes}\ \emph {et~al.}(2017)\citenamefont {Faes} \emph {et~al.}}]{faes2017multiscale}%
  \BibitemOpen
  \bibfield  {author} {\bibinfo {author} {\bibfnamefont {L.}~\bibnamefont {Faes}} \emph {et~al.},\ }\bibfield  {title} {\bibinfo {title} {Multiscale information decomposition: Exact computation for multivariate gaussian processes},\ }\href@noop {} {\bibfield  {journal} {\bibinfo  {journal} {Entropy}\ }\textbf {\bibinfo {volume} {19}},\ \bibinfo {pages} {408} (\bibinfo {year} {2017})}\BibitemShut {NoStop}%
\bibitem [{\citenamefont {Faes}\ \emph {et~al.}(2013)\citenamefont {Faes}, \citenamefont {Porta}, \citenamefont {Rossato}, \citenamefont {Adami}, \citenamefont {Tonon}, \citenamefont {Corica},\ and\ \citenamefont {Nollo}}]{faes2013investigating}%
  \BibitemOpen
  \bibfield  {author} {\bibinfo {author} {\bibfnamefont {L.}~\bibnamefont {Faes}}, \bibinfo {author} {\bibfnamefont {A.}~\bibnamefont {Porta}}, \bibinfo {author} {\bibfnamefont {G.}~\bibnamefont {Rossato}}, \bibinfo {author} {\bibfnamefont {A.}~\bibnamefont {Adami}}, \bibinfo {author} {\bibfnamefont {D.}~\bibnamefont {Tonon}}, \bibinfo {author} {\bibfnamefont {A.}~\bibnamefont {Corica}},\ and\ \bibinfo {author} {\bibfnamefont {G.}~\bibnamefont {Nollo}},\ }\bibfield  {title} {\bibinfo {title} {Investigating the mechanisms of cardiovascular and cerebrovascular regulation in orthostatic syncope through an information decomposition strategy},\ }\href@noop {} {\bibfield  {journal} {\bibinfo  {journal} {Autonomic Neuroscience}\ }\textbf {\bibinfo {volume} {178}},\ \bibinfo {pages} {76} (\bibinfo {year} {2013})}\BibitemShut {NoStop}%
\bibitem [{\citenamefont {Bari}\ \emph {et~al.}(2016{\natexlab{a}})\citenamefont {Bari}, \citenamefont {Marchi}, \citenamefont {Maria}, \citenamefont {Rossato}, \citenamefont {Nollo}, \citenamefont {Faes},\ and\ \citenamefont {Porta}}]{bari2016}%
  \BibitemOpen
  \bibfield  {author} {\bibinfo {author} {\bibfnamefont {V.}~\bibnamefont {Bari}}, \bibinfo {author} {\bibfnamefont {A.}~\bibnamefont {Marchi}}, \bibinfo {author} {\bibfnamefont {B.~D.}\ \bibnamefont {Maria}}, \bibinfo {author} {\bibfnamefont {G.}~\bibnamefont {Rossato}}, \bibinfo {author} {\bibfnamefont {G.}~\bibnamefont {Nollo}}, \bibinfo {author} {\bibfnamefont {L.}~\bibnamefont {Faes}},\ and\ \bibinfo {author} {\bibfnamefont {A.}~\bibnamefont {Porta}},\ }\bibfield  {title} {\bibinfo {title} {Nonlinear effects of respiration on the crosstalk between cardiovascular and cerebrovascular control systems},\ }\href@noop {} {\bibfield  {journal} {\bibinfo  {journal} {Phil. Trans. Royal Soc. A}\ }\textbf {\bibinfo {volume} {374}},\ \bibinfo {pages} {20150179} (\bibinfo {year} {2016}{\natexlab{a}})}\BibitemShut {NoStop}%
\bibitem [{\citenamefont {Pagani}\ \emph {et~al.}(1986)\citenamefont {Pagani}, \citenamefont {Lombardi}, \citenamefont {Guzzetti}, \citenamefont {Rimoldi}, \citenamefont {Furlan}, \citenamefont {Pizzinelli}, \citenamefont {Sandrone}, \citenamefont {Malfatto}, \citenamefont {Dell'Orto},\ and\ \citenamefont {Piccaluga}}]{pagani1986power}%
  \BibitemOpen
  \bibfield  {author} {\bibinfo {author} {\bibfnamefont {M.}~\bibnamefont {Pagani}}, \bibinfo {author} {\bibfnamefont {F.}~\bibnamefont {Lombardi}}, \bibinfo {author} {\bibfnamefont {S.}~\bibnamefont {Guzzetti}}, \bibinfo {author} {\bibfnamefont {O.}~\bibnamefont {Rimoldi}}, \bibinfo {author} {\bibfnamefont {R.}~\bibnamefont {Furlan}}, \bibinfo {author} {\bibfnamefont {P.}~\bibnamefont {Pizzinelli}}, \bibinfo {author} {\bibfnamefont {G.}~\bibnamefont {Sandrone}}, \bibinfo {author} {\bibfnamefont {G.}~\bibnamefont {Malfatto}}, \bibinfo {author} {\bibfnamefont {S.}~\bibnamefont {Dell'Orto}},\ and\ \bibinfo {author} {\bibfnamefont {E.}~\bibnamefont {Piccaluga}},\ }\bibfield  {title} {\bibinfo {title} {Power spectral analysis of heart rate and arterial pressure variabilities as a marker of sympatho-vagal interaction in man and conscious dog.},\ }\href@noop {} {\bibfield  {journal} {\bibinfo  {journal} {Circulation research}\ }\textbf {\bibinfo {volume} {59}},\ \bibinfo {pages} {178} (\bibinfo {year}
  {1986})}\BibitemShut {NoStop}%
\bibitem [{\citenamefont {Montano}\ \emph {et~al.}(1994)\citenamefont {Montano}, \citenamefont {Ruscone}, \citenamefont {Porta}, \citenamefont {Lombardi}, \citenamefont {Pagani},\ and\ \citenamefont {Malliani}}]{montano1994power}%
  \BibitemOpen
  \bibfield  {author} {\bibinfo {author} {\bibfnamefont {N.}~\bibnamefont {Montano}}, \bibinfo {author} {\bibfnamefont {T.~G.}\ \bibnamefont {Ruscone}}, \bibinfo {author} {\bibfnamefont {A.}~\bibnamefont {Porta}}, \bibinfo {author} {\bibfnamefont {F.}~\bibnamefont {Lombardi}}, \bibinfo {author} {\bibfnamefont {M.}~\bibnamefont {Pagani}},\ and\ \bibinfo {author} {\bibfnamefont {A.}~\bibnamefont {Malliani}},\ }\bibfield  {title} {\bibinfo {title} {Power spectrum analysis of heart rate variability to assess the changes in sympathovagal balance during graded orthostatic tilt.},\ }\href@noop {} {\bibfield  {journal} {\bibinfo  {journal} {Circulation}\ }\textbf {\bibinfo {volume} {90}},\ \bibinfo {pages} {1826} (\bibinfo {year} {1994})}\BibitemShut {NoStop}%
\bibitem [{\citenamefont {Cooley}\ \emph {et~al.}(1998)\citenamefont {Cooley}, \citenamefont {Montano}, \citenamefont {Cogliati}, \citenamefont {Van~de Borne}, \citenamefont {Richenbacher}, \citenamefont {Oren},\ and\ \citenamefont {Somers}}]{cooley1998evidence}%
  \BibitemOpen
  \bibfield  {author} {\bibinfo {author} {\bibfnamefont {R.~L.}\ \bibnamefont {Cooley}}, \bibinfo {author} {\bibfnamefont {N.}~\bibnamefont {Montano}}, \bibinfo {author} {\bibfnamefont {C.}~\bibnamefont {Cogliati}}, \bibinfo {author} {\bibfnamefont {P.}~\bibnamefont {Van~de Borne}}, \bibinfo {author} {\bibfnamefont {W.}~\bibnamefont {Richenbacher}}, \bibinfo {author} {\bibfnamefont {R.}~\bibnamefont {Oren}},\ and\ \bibinfo {author} {\bibfnamefont {V.~K.}\ \bibnamefont {Somers}},\ }\bibfield  {title} {\bibinfo {title} {Evidence for a central origin of the low-frequency oscillation in rr-interval variability},\ }\href@noop {} {\bibfield  {journal} {\bibinfo  {journal} {Circulation}\ }\textbf {\bibinfo {volume} {98}},\ \bibinfo {pages} {556} (\bibinfo {year} {1998})}\BibitemShut {NoStop}%
\bibitem [{\citenamefont {Stefanovska}(2002)}]{stefanovska2002cardiorespiratory}%
  \BibitemOpen
  \bibfield  {author} {\bibinfo {author} {\bibfnamefont {A.}~\bibnamefont {Stefanovska}},\ }\bibfield  {title} {\bibinfo {title} {Cardiorespiratory interactions},\ }\href@noop {} {\bibfield  {journal} {\bibinfo  {journal} {Nonlinear Phenomena in complex systems}\ }\textbf {\bibinfo {volume} {5}},\ \bibinfo {pages} {462} (\bibinfo {year} {2002})}\BibitemShut {NoStop}%
\bibitem [{\citenamefont {Zhang}\ \emph {et~al.}(2000)\citenamefont {Zhang}, \citenamefont {Zuckerman},\ and\ \citenamefont {Levine}}]{zhang2000spontaneous}%
  \BibitemOpen
  \bibfield  {author} {\bibinfo {author} {\bibfnamefont {R.}~\bibnamefont {Zhang}}, \bibinfo {author} {\bibfnamefont {J.~H.}\ \bibnamefont {Zuckerman}},\ and\ \bibinfo {author} {\bibfnamefont {B.~D.}\ \bibnamefont {Levine}},\ }\bibfield  {title} {\bibinfo {title} {Spontaneous fluctuations in cerebral blood flow: insights from extended-duration recordings in humans},\ }\href@noop {} {\bibfield  {journal} {\bibinfo  {journal} {American Journal of Physiology-Heart and Circulatory Physiology}\ }\textbf {\bibinfo {volume} {278}},\ \bibinfo {pages} {H1848} (\bibinfo {year} {2000})}\BibitemShut {NoStop}%
\bibitem [{\citenamefont {Bari}\ \emph {et~al.}(2016{\natexlab{b}})\citenamefont {Bari}, \citenamefont {Marchi}, \citenamefont {De~Maria}, \citenamefont {Rossato}, \citenamefont {Nollo}, \citenamefont {Faes},\ and\ \citenamefont {Porta}}]{bari2016nonlinear}%
  \BibitemOpen
  \bibfield  {author} {\bibinfo {author} {\bibfnamefont {V.}~\bibnamefont {Bari}}, \bibinfo {author} {\bibfnamefont {A.}~\bibnamefont {Marchi}}, \bibinfo {author} {\bibfnamefont {B.}~\bibnamefont {De~Maria}}, \bibinfo {author} {\bibfnamefont {G.}~\bibnamefont {Rossato}}, \bibinfo {author} {\bibfnamefont {G.}~\bibnamefont {Nollo}}, \bibinfo {author} {\bibfnamefont {L.}~\bibnamefont {Faes}},\ and\ \bibinfo {author} {\bibfnamefont {A.}~\bibnamefont {Porta}},\ }\bibfield  {title} {\bibinfo {title} {Nonlinear effects of respiration on the crosstalk between cardiovascular and cerebrovascular control systems},\ }\href@noop {} {\bibfield  {journal} {\bibinfo  {journal} {Philosophical Transactions of the Royal Society A: Mathematical, Physical and Engineering Sciences}\ }\textbf {\bibinfo {volume} {374}},\ \bibinfo {pages} {20150179} (\bibinfo {year} {2016}{\natexlab{b}})}\BibitemShut {NoStop}%
\bibitem [{\citenamefont {Faes}\ \emph {et~al.}(2022{\natexlab{b}})\citenamefont {Faes}, \citenamefont {Mijatovic}, \citenamefont {Sparacino}, \citenamefont {Pernice}, \citenamefont {Antonacci}, \citenamefont {Porta},\ and\ \citenamefont {Stramaglia}}]{faes2022quantifying}%
  \BibitemOpen
  \bibfield  {author} {\bibinfo {author} {\bibfnamefont {L.}~\bibnamefont {Faes}}, \bibinfo {author} {\bibfnamefont {G.}~\bibnamefont {Mijatovic}}, \bibinfo {author} {\bibfnamefont {L.}~\bibnamefont {Sparacino}}, \bibinfo {author} {\bibfnamefont {R.}~\bibnamefont {Pernice}}, \bibinfo {author} {\bibfnamefont {Y.}~\bibnamefont {Antonacci}}, \bibinfo {author} {\bibfnamefont {A.}~\bibnamefont {Porta}},\ and\ \bibinfo {author} {\bibfnamefont {S.}~\bibnamefont {Stramaglia}},\ }\bibfield  {title} {\bibinfo {title} {Quantifying high-order interactions in cardiovascular and cerebrovascular networks},\ }in\ \href@noop {} {\emph {\bibinfo {booktitle} {2022 12th Conference of the European Study Group on Cardiovascular Oscillations (ESGCO)}}}\ (\bibinfo {organization} {IEEE},\ \bibinfo {year} {2022})\ pp.\ \bibinfo {pages} {1--2}\BibitemShut {NoStop}%
\bibitem [{\citenamefont {Sparacino}\ \emph {et~al.}(2022)\citenamefont {Sparacino}, \citenamefont {Antonacci}, \citenamefont {Marinazzo}, \citenamefont {Stramaglia},\ and\ \citenamefont {Faes}}]{sparacino2022quantifying}%
  \BibitemOpen
  \bibfield  {author} {\bibinfo {author} {\bibfnamefont {L.}~\bibnamefont {Sparacino}}, \bibinfo {author} {\bibfnamefont {Y.}~\bibnamefont {Antonacci}}, \bibinfo {author} {\bibfnamefont {D.}~\bibnamefont {Marinazzo}}, \bibinfo {author} {\bibfnamefont {S.}~\bibnamefont {Stramaglia}},\ and\ \bibinfo {author} {\bibfnamefont {L.}~\bibnamefont {Faes}},\ }\bibfield  {title} {\bibinfo {title} {Quantifying high-order interactions in complex physiological networks: A frequency-specific approach},\ }in\ \href@noop {} {\emph {\bibinfo {booktitle} {International Conference on Complex Networks and Their Applications}}}\ (\bibinfo {organization} {Springer},\ \bibinfo {year} {2022})\ pp.\ \bibinfo {pages} {301--309}\BibitemShut {NoStop}%
\bibitem [{\citenamefont {Scagliarini}\ \emph {et~al.}(2023)\citenamefont {Scagliarini}, \citenamefont {Sparacino}, \citenamefont {Faes}, \citenamefont {Marinazzo},\ and\ \citenamefont {Stramaglia}}]{scagliarini3gradients}%
  \BibitemOpen
  \bibfield  {author} {\bibinfo {author} {\bibfnamefont {T.}~\bibnamefont {Scagliarini}}, \bibinfo {author} {\bibfnamefont {L.}~\bibnamefont {Sparacino}}, \bibinfo {author} {\bibfnamefont {L.}~\bibnamefont {Faes}}, \bibinfo {author} {\bibfnamefont {D.}~\bibnamefont {Marinazzo}},\ and\ \bibinfo {author} {\bibfnamefont {S.}~\bibnamefont {Stramaglia}},\ }\bibfield  {title} {\bibinfo {title} {Gradients of o-information highlight synergy and redundancy in physiological applications},\ }\href@noop {} {\bibfield  {journal} {\bibinfo  {journal} {Frontiers in Network Physiology}\ }\textbf {\bibinfo {volume} {3}},\ \bibinfo {pages} {1335808} (\bibinfo {year} {2023})}\BibitemShut {NoStop}%
\bibitem [{\citenamefont {Schalk}\ \emph {et~al.}(2004)\citenamefont {Schalk}, \citenamefont {McFarland}, \citenamefont {Hinterberger}, \citenamefont {Birbaumer},\ and\ \citenamefont {Wolpaw}}]{schalk2004bci2000}%
  \BibitemOpen
  \bibfield  {author} {\bibinfo {author} {\bibfnamefont {G.}~\bibnamefont {Schalk}}, \bibinfo {author} {\bibfnamefont {D.~J.}\ \bibnamefont {McFarland}}, \bibinfo {author} {\bibfnamefont {T.}~\bibnamefont {Hinterberger}}, \bibinfo {author} {\bibfnamefont {N.}~\bibnamefont {Birbaumer}},\ and\ \bibinfo {author} {\bibfnamefont {J.~R.}\ \bibnamefont {Wolpaw}},\ }\bibfield  {title} {\bibinfo {title} {Bci2000: a general-purpose brain-computer interface (bci) system},\ }\href@noop {} {\bibfield  {journal} {\bibinfo  {journal} {IEEE Transactions on biomedical engineering}\ }\textbf {\bibinfo {volume} {51}},\ \bibinfo {pages} {1034} (\bibinfo {year} {2004})}\BibitemShut {NoStop}%
\bibitem [{\citenamefont {Goldberger}\ \emph {et~al.}(2000)\citenamefont {Goldberger}, \citenamefont {Amaral}, \citenamefont {Glass}, \citenamefont {Hausdorff}, \citenamefont {Ivanov}, \citenamefont {Mark}, \citenamefont {Mietus}, \citenamefont {Moody}, \citenamefont {Peng},\ and\ \citenamefont {Stanley}}]{goldberger2000physiobank}%
  \BibitemOpen
  \bibfield  {author} {\bibinfo {author} {\bibfnamefont {A.~L.}\ \bibnamefont {Goldberger}}, \bibinfo {author} {\bibfnamefont {L.~A.}\ \bibnamefont {Amaral}}, \bibinfo {author} {\bibfnamefont {L.}~\bibnamefont {Glass}}, \bibinfo {author} {\bibfnamefont {J.~M.}\ \bibnamefont {Hausdorff}}, \bibinfo {author} {\bibfnamefont {P.~C.}\ \bibnamefont {Ivanov}}, \bibinfo {author} {\bibfnamefont {R.~G.}\ \bibnamefont {Mark}}, \bibinfo {author} {\bibfnamefont {J.~E.}\ \bibnamefont {Mietus}}, \bibinfo {author} {\bibfnamefont {G.~B.}\ \bibnamefont {Moody}}, \bibinfo {author} {\bibfnamefont {C.-K.}\ \bibnamefont {Peng}},\ and\ \bibinfo {author} {\bibfnamefont {H.~E.}\ \bibnamefont {Stanley}},\ }\bibfield  {title} {\bibinfo {title} {Physiobank, physiotoolkit, and physionet: components of a new research resource for complex physiologic signals},\ }\href@noop {} {\bibfield  {journal} {\bibinfo  {journal} {circulation}\ }\textbf {\bibinfo {volume} {101}},\ \bibinfo {pages} {e215} (\bibinfo {year} {2000})}\BibitemShut {NoStop}%
\bibitem [{\citenamefont {Cona}\ \emph {et~al.}(2009)\citenamefont {Cona}, \citenamefont {Zavaglia}, \citenamefont {Astolfi}, \citenamefont {Babiloni}, \citenamefont {Ursino} \emph {et~al.}}]{cona2009changes}%
  \BibitemOpen
  \bibfield  {author} {\bibinfo {author} {\bibfnamefont {F.}~\bibnamefont {Cona}}, \bibinfo {author} {\bibfnamefont {M.}~\bibnamefont {Zavaglia}}, \bibinfo {author} {\bibfnamefont {L.}~\bibnamefont {Astolfi}}, \bibinfo {author} {\bibfnamefont {F.}~\bibnamefont {Babiloni}}, \bibinfo {author} {\bibfnamefont {M.}~\bibnamefont {Ursino}}, \emph {et~al.},\ }\bibfield  {title} {\bibinfo {title} {Changes in eeg power spectral density and cortical connectivity in healthy and tetraplegic patients during a motor imagery task},\ }\href@noop {} {\bibfield  {journal} {\bibinfo  {journal} {Computational intelligence and neuroscience}\ }\textbf {\bibinfo {volume} {2009}} (\bibinfo {year} {2009})}\BibitemShut {NoStop}%
\bibitem [{\citenamefont {Pirovano}\ \emph {et~al.}(2022)\citenamefont {Pirovano}, \citenamefont {Mastropietro}, \citenamefont {Antonacci}, \citenamefont {Bar{\`a}}, \citenamefont {Guanziroli}, \citenamefont {Molteni}, \citenamefont {Faes},\ and\ \citenamefont {Rizzo}}]{pirovano2022resting}%
  \BibitemOpen
  \bibfield  {author} {\bibinfo {author} {\bibfnamefont {I.}~\bibnamefont {Pirovano}}, \bibinfo {author} {\bibfnamefont {A.}~\bibnamefont {Mastropietro}}, \bibinfo {author} {\bibfnamefont {Y.}~\bibnamefont {Antonacci}}, \bibinfo {author} {\bibfnamefont {C.}~\bibnamefont {Bar{\`a}}}, \bibinfo {author} {\bibfnamefont {E.}~\bibnamefont {Guanziroli}}, \bibinfo {author} {\bibfnamefont {F.}~\bibnamefont {Molteni}}, \bibinfo {author} {\bibfnamefont {L.}~\bibnamefont {Faes}},\ and\ \bibinfo {author} {\bibfnamefont {G.}~\bibnamefont {Rizzo}},\ }\bibfield  {title} {\bibinfo {title} {Resting state eeg directed functional connectivity unveils changes in motor network organization in subacute stroke patients after rehabilitation},\ }\href@noop {} {\bibfield  {journal} {\bibinfo  {journal} {Frontiers in Physiology}\ }\textbf {\bibinfo {volume} {13}},\ \bibinfo {pages} {862207} (\bibinfo {year} {2022})}\BibitemShut {NoStop}%
\bibitem [{\citenamefont {Grefkes}\ \emph {et~al.}(2008)\citenamefont {Grefkes}, \citenamefont {Eickhoff}, \citenamefont {Nowak}, \citenamefont {Dafotakis},\ and\ \citenamefont {Fink}}]{grefkes2008dynamic}%
  \BibitemOpen
  \bibfield  {author} {\bibinfo {author} {\bibfnamefont {C.}~\bibnamefont {Grefkes}}, \bibinfo {author} {\bibfnamefont {S.~B.}\ \bibnamefont {Eickhoff}}, \bibinfo {author} {\bibfnamefont {D.~A.}\ \bibnamefont {Nowak}}, \bibinfo {author} {\bibfnamefont {M.}~\bibnamefont {Dafotakis}},\ and\ \bibinfo {author} {\bibfnamefont {G.~R.}\ \bibnamefont {Fink}},\ }\bibfield  {title} {\bibinfo {title} {Dynamic intra-and interhemispheric interactions during unilateral and bilateral hand movements assessed with fmri and dcm},\ }\href@noop {} {\bibfield  {journal} {\bibinfo  {journal} {Neuroimage}\ }\textbf {\bibinfo {volume} {41}},\ \bibinfo {pages} {1382} (\bibinfo {year} {2008})}\BibitemShut {NoStop}%
\bibitem [{\citenamefont {Pirovano}\ \emph {et~al.}(2023)\citenamefont {Pirovano}, \citenamefont {Antonacci}, \citenamefont {Mastropietro}, \citenamefont {Bar{\`a}}, \citenamefont {Sparacino}, \citenamefont {Guanziroli}, \citenamefont {Molteni}, \citenamefont {Tettamanti}, \citenamefont {Faes},\ and\ \citenamefont {Rizzo}}]{pirovano2023rehabilitation}%
  \BibitemOpen
  \bibfield  {author} {\bibinfo {author} {\bibfnamefont {I.}~\bibnamefont {Pirovano}}, \bibinfo {author} {\bibfnamefont {Y.}~\bibnamefont {Antonacci}}, \bibinfo {author} {\bibfnamefont {A.}~\bibnamefont {Mastropietro}}, \bibinfo {author} {\bibfnamefont {C.}~\bibnamefont {Bar{\`a}}}, \bibinfo {author} {\bibfnamefont {L.}~\bibnamefont {Sparacino}}, \bibinfo {author} {\bibfnamefont {E.}~\bibnamefont {Guanziroli}}, \bibinfo {author} {\bibfnamefont {F.}~\bibnamefont {Molteni}}, \bibinfo {author} {\bibfnamefont {M.}~\bibnamefont {Tettamanti}}, \bibinfo {author} {\bibfnamefont {L.}~\bibnamefont {Faes}},\ and\ \bibinfo {author} {\bibfnamefont {G.}~\bibnamefont {Rizzo}},\ }\bibfield  {title} {\bibinfo {title} {Rehabilitation modulates high-order interactions among large-scale brain networks in subacute stroke},\ }\href@noop {} {\bibfield  {journal} {\bibinfo  {journal} {IEEE Transactions on Neural Systems and Rehabilitation Engineering}\ }\textbf {\bibinfo {volume} {31}},\ \bibinfo {pages} {4549} (\bibinfo {year}
  {2023})}\BibitemShut {NoStop}%
\bibitem [{\citenamefont {Van~de Steen}\ \emph {et~al.}(2019)\citenamefont {Van~de Steen}, \citenamefont {Faes}, \citenamefont {Karahan}, \citenamefont {Songsiri}, \citenamefont {Valdes-Sosa},\ and\ \citenamefont {Marinazzo}}]{van2019critical}%
  \BibitemOpen
  \bibfield  {author} {\bibinfo {author} {\bibfnamefont {F.}~\bibnamefont {Van~de Steen}}, \bibinfo {author} {\bibfnamefont {L.}~\bibnamefont {Faes}}, \bibinfo {author} {\bibfnamefont {E.}~\bibnamefont {Karahan}}, \bibinfo {author} {\bibfnamefont {J.}~\bibnamefont {Songsiri}}, \bibinfo {author} {\bibfnamefont {P.~A.}\ \bibnamefont {Valdes-Sosa}},\ and\ \bibinfo {author} {\bibfnamefont {D.}~\bibnamefont {Marinazzo}},\ }\bibfield  {title} {\bibinfo {title} {Critical comments on eeg sensor space dynamical connectivity analysis},\ }\href@noop {} {\bibfield  {journal} {\bibinfo  {journal} {Brain topography}\ }\textbf {\bibinfo {volume} {32}},\ \bibinfo {pages} {643} (\bibinfo {year} {2019})}\BibitemShut {NoStop}%
\end{thebibliography}%

\end{document}